\documentclass[a4paper,UKenglish,autoref]{lipics-v2021}
\pdfoutput=1
\graphicspath{{./figures/}}
\usepackage{tikz}
\usetikzlibrary{arrows.meta, positioning, shapes.geometric, fit, backgrounds}
\usepackage{float}
\usepackage{adjustbox}
\usepackage{tcolorbox}
\usepackage{booktabs}
\usepackage{makecell}
\usepackage{multirow}
\usepackage{rotating}
\usepackage{array}
\usepackage{caption}
\usepackage{threeparttable}
\tcbuselibrary{skins}
\usepackage{graphicx}
\usepackage{placeins}
\usepackage[normalem]{ulem}
\usepackage[table]{xcolor}
\definecolor{cliffNeg}{HTML}{F5F5F5}
\definecolor{cliffSmall}{HTML}{FFE5B4}
\definecolor{cliffMed}{HTML}{FFB347}
\definecolor{cliffLarge}{HTML}{E74C3C}
\newcommand{\cN}{\cellcolor{cliffNeg}N}
\newcommand{\cS}{\cellcolor{cliffSmall}S}
\newcommand{\cM}{\cellcolor{cliffMed}M}
\newcommand{\cL}{\cellcolor{cliffLarge}L}
\newtcolorbox{finding}[1]{
  colback=blue!4, colframe=blue!55!black, coltitle=white,
  fonttitle=\bfseries\small, fontupper=\small,
  title={#1},
  boxrule=0.5pt, arc=1pt, left=4pt, right=4pt, top=3pt, bottom=3pt,
  before skip=4pt, after skip=4pt,
}

\newtcolorbox{rqsummary}[1]{
  colback=gray!6, colframe=black!60, coltitle=black,
  fonttitle=\bfseries\small, fontupper=\small,
  title={RQ#1 summary},
  boxrule=0.5pt, arc=1pt, left=4pt, right=4pt, top=3pt, bottom=3pt,
  before skip=6pt, after skip=6pt,
}

\title{How Developers Use Relation Chains in Gerrit-Based Review Ecosystems: An Empirical Study Across Three Open-Source Ecosystems}

\author{Ahmed Belhouchette}{
ENSI, Manouba University, Tunisia
}{ahmed.belhouchette@ensi-uma.tn}{}{}

\author{Moataz Chouchen}{
Concordia University, Montreal, QC, Canada
}{moataz.chouchen@concordia.ca}{}{}

\author{Marouene Chaieb}{
ENSI, Manouba University, Tunisia\\
LARODEC Laboratory, ISG-Tunis, Tunis University, Tunisia
}{marouene.chaieb@ensi-uma.tn}{}{}

\author{Mohammad Hamdaqa}{
Polytechnique Montréal, Montreal, QC, Canada
}{mhamdaqa@polymtl.ca}{}{}

\author{Abdelwahab Hamou-Lhadj}{
Concordia University, Montreal, QC, Canada
}{wahab.hamou-lhadj@concordia.ca}{}{}

\authorrunning{A. Belhouchette, M. Chouchen, M. Chaieb, M. Hamdaqa, and A. Hamou-Lhadj}

\Copyright{
Ahmed Belhouchette,
Moataz Chouchen,
Marouene Chaieb,
Mohammad Hamdaqa,
and Abdelwahab Hamou-Lhadj
}

\ccsdesc[500]{Software and its engineering~Software configuration management and version control systems}
\ccsdesc[300]{Software and its engineering~Collaboration in software development}
\ccsdesc[300]{Software and its engineering~Software verification and validation}

\keywords{Code review, relation chains, stacked changes, dependent patches, empirical software engineering, Gerrit}

\begin{document}

\maketitle

\begin{abstract}
\noindent\textbf{Background.}
Developers increasingly coordinate dependent review workflows by submitting sequences of related changes rather than monolithic ones. In Gerrit, these dependencies form \emph{relation chains}: structured review units that link changes together. As chains become more common, they shape review activities through synchronization overhead, CI amplification, and merge-ordering constraints.

\noindent\textbf{Aim.}
We investigate how developers adopt relation chains and how these dependency structures influence review dynamics and outcomes.

\noindent\textbf{Method.}
We analyze 29{,}580 relation chains from 15~repositories across three Gerrit ecosystems (OpenStack, Wikimedia, ONAP), comprising 401{,}256 changes, using Mann--Kendall trend tests, Mann--Whitney with Cliff's $\delta$ for chain-vs-solo comparison, and Spearman correlations for base--descendant dependency.

\noindent\textbf{Results.}
Chain prevalence ranges from 5\% to 49\% across projects, increasing in 14 of 15. Chain changes take a median of 2.6$\times$ longer to merge than size-matched solo changes, with the gap widening for very large changes. Review effort propagates through dependency-linked review workflows: base-change review activity co-varies with descendant review activity ($\rho = 0.43$--$0.61$ in 14--15/15~projects), and 33.5\% of chain members undergo structural evolution during review.

\noindent\textbf{Conclusions.}
Relation chains operate as durable, ecosystem-shaped coordination units with internal structure that change-centric analyses cannot capture. Future review analytics, reviewer-assignment systems, and AI-assisted review tools should reason over chains rather than isolated changes.
\end{abstract}

\section{Introduction}
\label{sec:introduction}

Modern Code Review (MCR) is used to detect defects, discuss design decisions, transfer knowledge, and enforce coding conventions~\cite{bacchelli2013expectations,rigby2013convergent,mcintosh2016empirical,kononenko2016codereview,macleod2017code}. Large engineering tasks such as refactorings, migrations, and multi-component features are increasingly decomposed into sequences of dependent changes submitted as \emph{stacked changes}. In Gerrit, these structures are represented as \emph{relation chains}, a higher-level coordination abstraction in modern code-review workflows; similar abstractions exist in Phabricator, Graphite, GitHub, and \texttt{ghstack}~\cite{phabricator2024,graphite2024,github2026stacked,ghstack}.

\begin{figure}[!ht]
\centering
\begin{minipage}[c]{0.4\textwidth}
\centering
\includegraphics[width=\linewidth]{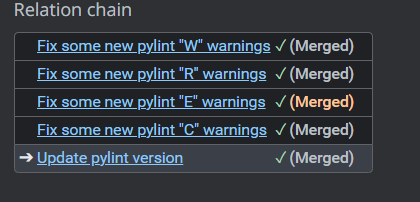}
{\small (a) Gerrit's \emph{Relation Chain} panel}
\end{minipage}\hfill
\begin{minipage}[c]{0.55\textwidth}
\centering
\resizebox{\linewidth}{!}{%
\begin{tikzpicture}[
  node distance=3mm and 3mm,
  every node/.style={font=\footnotesize},
  cbox/.style={
    draw, rounded corners=2pt, align=center,
    inner sep=4pt, minimum width=24mm, minimum height=8mm,
    fill=green!12, draw=green!50!black, line width=0.4pt
  },
  base/.style={
    cbox, fill=blue!12, draw=blue!60!black, line width=0.6pt
  },
  branch/.style={cbox},
  edge/.style={-{Stealth[length=2mm]}, thin, draw=black!60}
]
\node[base] (b) {\textbf{\#883464}\\Update pylint version};
\node[branch, below=of b, xshift=-30mm] (w) {\#883605\\Fix \texttt{W} warnings};
\node[branch, below=of b]                (r) {\#883606\\Fix \texttt{R} warnings};
\node[branch, below=of b, xshift=30mm]  (c) {\#883608\\Fix \texttt{C} warnings};
\node[branch, below=of w] (w2) {\#890244\\Fix \texttt{W}\\ \texttt{missing-timeout}};
\draw[edge] (b.south) -- (w.north);
\draw[edge] (b.south) -- (r.north);
\draw[edge] (b.south) -- (c.north);
\draw[edge] (w.south) -- (w2.north);
\end{tikzpicture}%
}\\[2pt]
{\small (b) Relation chain dependency graph.}
\end{minipage}
\caption{A branching relation chain in OpenStack Neutron (change~\#883464\protect\footnote{\url{https://review.opendev.org/c/openstack/neutron/+/883464}}). (a)~Gerrit's flat \emph{Relation Chain} panel. (b)~The same chain reconstructed from patchset-level parent SHAs: blue is the base, green is a descendant. All five members merged.}
\label{fig:gerrit-screenshot}
\end{figure}
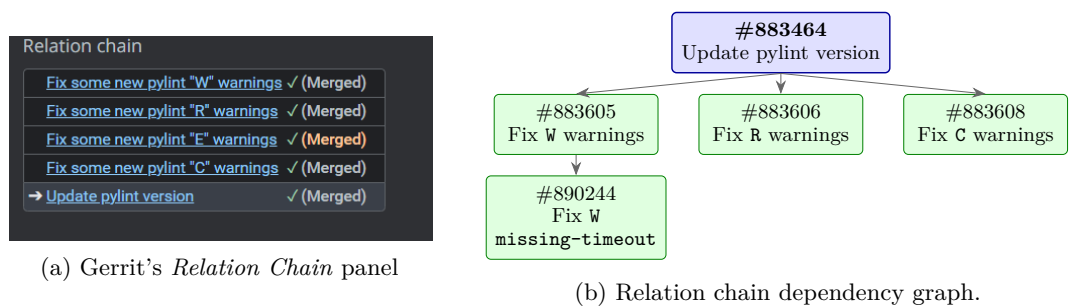

Following the Gerrit documentation~\cite{gerritdocs_relationchain}, we use the term \emph{relation chain} to denote a dependency-linked review structure in which multiple changes evolve simultaneously and the state of one review directly influences the validation, synchronization, and mergeability of dependent reviews. In this paper, relation chains refer specifically to Gerrit parent-SHA dependency structures reconstructed from patchset-level parent relationships; while similar workflows exist in GitHub stacked pull requests, Phabricator, and \texttt{ghstack}, our operationalization and findings are specific to Gerrit-based review ecosystems. Relation chains let reviewers reason about large engineering efforts as coordinated structures rather than isolated changes. Figure~\ref{fig:gerrit-screenshot} shows an OpenStack Neutron example: a \texttt{pylint} migration\footnote{\url{https://review.opendev.org/c/openstack/neutron/+/883464}} decomposed into descendants addressing different warning categories. The same dependency that enables this coordination can also become a source of friction: updates to one chain member can cascade across descendants through forced rebasing, repeated CI validation, and delayed merge propagation. On ONAP \texttt{oom} change~100114\footnote{\url{https://gerrit.onap.org/r/c/oom/+/100114}}, a reviewer warned the author that ``\emph{as parent is abandoned, we cannot merge\ldots please rebase on master},'' surfacing parent abandonment, merge blocking, and forced rebasing as everyday chain-coordination costs. Understanding how such dependency structures influence review processes matters for review analytics, prioritization, and tooling~\cite{thongtanunam2017review}.

Despite their growing adoption, prior research has overlooked relation chains as first-class review artifacts. Existing work captures only fragments: studies of cross-component dependencies~\cite{arabat2024cross} model pairwise relationships between repositories rather than the chain that links a sequence of changes within a single review; \texttt{Depends-On}/\texttt{Needed-By} annotation analyses~\cite{arabat2025ml} require the dependency to be declared in the commit message, missing chains created implicitly through parent-SHA links; and patch-linkage research~\cite{hirao2019linkage,wang2021linkage,wang2022crosspatch} traces references between review discussions but not the structural dependency graph that determines merge order. Properties that emerge only when a change is read together with its in-review parents and children (friction propagation from base to descendant, mid-review restructuring of the dependency graph, position-dependent review effort) fall outside what existing methods can observe. These workflow effects remain largely invisible when reviews are analyzed as isolated artifacts. The literature has no empirical characterisation of relation chains as the unit through which developers coordinate dependent review work at scale.

To address this gap, we present an empirical characterisation of relation chains as first-class review artifacts. We analyze 401{,}256 reviewed changes and 29{,}580 chains across three Gerrit ecosystems (OpenStack, Wikimedia, ONAP), selected to span different development cultures, CI-gating regimes, and community sizes. The study combines three lenses: temporal-trend analysis of chain adoption, stratified comparison of chain members against size-matched solo changes, and dependency-aware correlation analysis between base changes and their descendants. Together they let us characterise chains both as a structural feature of the review graph and as a coordination mechanism whose dynamics shape review outcomes that change-level studies cannot observe. Because our analyses are observational, we interpret the reported effects as associations rather than causal consequences of chain membership.

Our results converge on a single characterisation: relation chains are durable, ecosystem-shaped, internally-structured coordination units whose review dynamics differ from those of isolated changes. Each of the four findings below contributes evidence along one of these dimensions. First, chains are \emph{prevalent and growing}: prevalence ranges from 5\% to 49\% and increases in 10 of 15 projects, so dependent-change workflows are not a niche pattern. Second, chains carry \emph{review costs not visible at the change level}: members take a median of $2.6\times$ longer to merge than size-matched solo changes, with overhead concentrating in very large changes (Cliff's~$\delta=0.49$), and CI activity is amplified five- to twenty-fold under strict gating. Third, chains have \emph{internal structure that shapes review outcomes}: middle-of-chain members absorb the highest review effort, and base-change review activity co-varies with descendant activity ($\rho = 0.43$--$0.61$ in 14--15 of 15~projects), a regularity we call the \emph{foundation effect}. Fourth, chains are \emph{not static}: 33.5\% of members undergo a structural change in their parent SHA before merging, and chains can persist across multi-year intermittent timespans, with the longest merge gap reaching 2.85 years on a chain in OpenStack \texttt{Neutron}. These results call into question the default assumption underlying current review tooling, which processes each change independently~\cite{thongtanunam2015whoreview,mirsaeedi2020mitigating,li2022automating}, and motivate chain-aware review analytics, reviewer-assignment systems that prioritise members whose merge unblocks the largest downstream subgraph, and AI-assisted review agents that condition on chain position and base-change review state. 
\section{Background: Modern Code Review, Gerrit, and Relation Chains}
\label{sec:related-work}

Modern Code Review (MCR) is a tool-supported form of peer review used in industrial and open-source development~\cite{bacchelli2013expectations,rigby2013convergent,macleod2017code}, supporting defect detection, design discussion, knowledge transfer, and coding-convention enforcement~\cite{mcintosh2016empirical,kononenko2016codereview,baysal2016investigating}. Smaller, focused changes are easier to review and more likely to be accepted~\cite{weissgerber2008small,ram2018reviewability}; larger engineering tasks therefore motivate developers to split work across related changes.

Gerrit is a web-based code review platform used by projects such as OpenStack, Android, Chromium, Wikimedia, and ONAP~\cite{gerritdocs2024,thongtanunam2017review}. Each review request is a \emph{change} identified by a \texttt{Change-Id} and refined through successive \emph{patchsets}. When a developer uploads a change whose parent commit corresponds to another change still under review, Gerrit links them as a parent--child dependency; a sequence of such linked changes forms a \emph{relation chain}~\cite{gerritdocs_relationchain}.

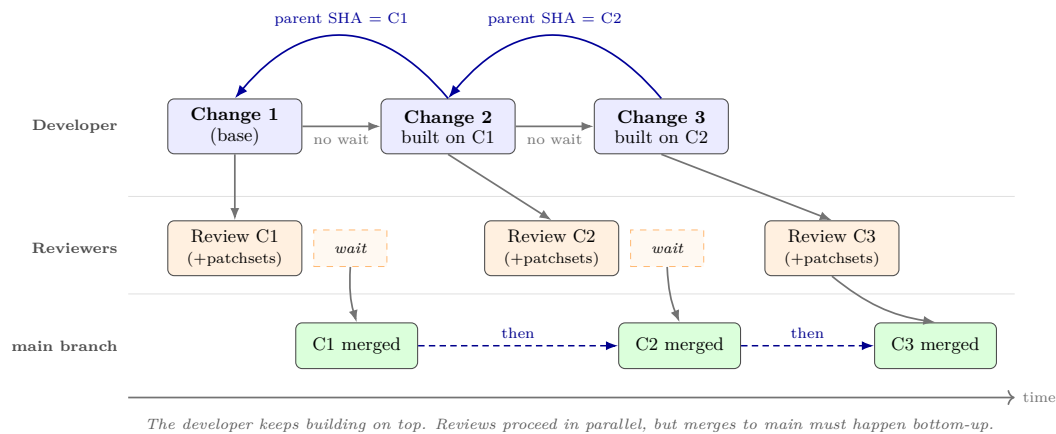
\begin{figure}[!ht]
\centering
\resizebox{\textwidth}{!}{%
\begin{tikzpicture}[
  font=\footnotesize,
  devbox/.style={rectangle, draw=black!70, rounded corners=3pt, minimum width=2.2cm, minimum height=0.9cm, align=center, fill=blue!8, line width=0.5pt},
  revbox/.style={rectangle, draw=black!70, rounded corners=3pt, minimum width=2.2cm, minimum height=0.9cm, align=center, fill=orange!12, line width=0.5pt},
  mergebox/.style={rectangle, draw=black!70, rounded corners=3pt, minimum width=2.0cm, minimum height=0.75cm, align=center, fill=green!14, line width=0.5pt},
  wait/.style={rectangle, draw=orange!55, dashed, fill=orange!5, minimum width=1.2cm, minimum height=0.6cm, align=center, font=\scriptsize\itshape, line width=0.4pt},
  lane/.style={font=\scriptsize\bfseries, anchor=east, text=black!70},
  flow/.style={-{Latex[length=2.0mm]}, thick, black!55},
  dep/.style={-{Latex[length=2.2mm]}, thick, blue!60!black, line width=0.8pt},
  mergedep/.style={-{Latex[length=2.0mm]}, densely dashed, thick, blue!55!black, line width=0.7pt},
]
\draw[black!12, line width=0.3pt] (-0.25, 2.45) -- (14.3, 2.45);
\draw[black!12, line width=0.3pt] (-0.25, 0.85) -- (14.3, 0.85);
\node[lane] at (-0.3, 3.6) {Developer};
\node[lane] at (-0.3, 1.6) {Reviewers};
\node[lane] at (-0.3, 0.0) {main branch};
\node[devbox] (C1) at (1.5, 3.6) {\textbf{Change 1}\\(base)};
\node[devbox] (C2) at (5.0, 3.6) {\textbf{Change 2}\\built on C1};
\node[devbox] (C3) at (8.5, 3.6) {\textbf{Change 3}\\built on C2};
\draw[flow] (C1.east) -- node[below, font=\scriptsize, text=black!55] {no wait} (C2.west);
\draw[flow] (C2.east) -- node[below, font=\scriptsize, text=black!55] {no wait} (C3.west);
\draw[dep] (C2.north) to[out=130, in=50, looseness=1.3]
  node[above, font=\scriptsize, text=blue!60!black, pos=0.5, yshift=1pt] {parent SHA $=$ C1}
  (C1.north);
\draw[dep] (C3.north) to[out=130, in=50, looseness=1.3]
  node[above, font=\scriptsize, text=blue!60!black, pos=0.5, yshift=1pt] {parent SHA $=$ C2}
  (C2.north);
\node[revbox] (R1) at (1.5, 1.6) {Review C1\\\scriptsize (+patchsets)};
\node[wait,   right=0.18cm of R1] (W1) {wait};
\node[revbox] (R2) at (6.7, 1.6) {Review C2\\\scriptsize (+patchsets)};
\node[wait,   right=0.18cm of R2] (W2) {wait};
\node[revbox] (R3) at (11.3, 1.6) {Review C3\\\scriptsize (+patchsets)};
\draw[flow] (C1.south) -- (R1.north);
\draw[flow] (C2.south) -- (R2.north);
\draw[flow] (C3.south) -- (R3.north);
\node[mergebox] (M1) at (3.5, 0.0) {C1 merged};
\node[mergebox] (M2) at (8.8, 0.0) {C2 merged};
\node[mergebox] (M3) at (13.0, 0.0) {C3 merged};
\draw[flow] (W1.south) to[bend right=10] (M1.north);
\draw[flow] (W2.south) to[bend right=10] (M2.north);
\draw[flow] (R3.south) to[bend right=15] (M3.north);
\draw[mergedep] (M1.east) -- node[above, font=\scriptsize, text=blue!55!black] {then} (M2.west);
\draw[mergedep] (M2.east) -- node[above, font=\scriptsize, text=blue!55!black] {then} (M3.west);
\draw[->, thick, black!55] (-0.25, -0.85) -- (14.3, -0.85) node[right, font=\scriptsize] {time};
\node[align=center, font=\scriptsize\itshape, text=black!65] at (7.0, -1.3)
  {The developer keeps building on top. Reviews proceed in parallel, but merges to main must happen bottom-up.};
\end{tikzpicture}%
}
\caption{Working with a relation chain in Gerrit. The developer builds C2 on C1 and C3 on C2 without waiting. Reviews proceed in parallel; merges into the main branch are bottom-up.}
\label{fig:gerrit-chain}
\end{figure}

Relation chains let developers continue dependent work without waiting for earlier changes to merge. The same workflow appears in Phabricator~\cite{phabricator2024}, Graphite~\cite{graphite2024}, Meta's \texttt{ghstack}~\cite{ghstack}, and GitHub's native stacked PRs~\cite{github2026stacked}; earlier work on Linux kernel patch series~\cite{rigby2008open} shows dependent review workflows predate modern platforms. Figure~\ref{fig:gerrit-chain} illustrates a typical workflow. Throughout this paper, \emph{base}, \emph{middle}, and \emph{top} denote the first, intermediate, and last members of a chain, and \emph{solo} denotes a change with no in-review parent or child.

\section{Related Work}

We position our study with respect to three lines of work: (i) empirical studies of modern code review that examine review outcomes at the change level, (ii) reviewer recommendation and review automation systems that consume change-level features, and (iii) change decomposition and dependent-review research that touches on multi-change workflows. For each, we summarise what the line has shown and clarify how our chain-level focus differs.

\subsection{Empirical Studies in Modern Code Review}
\label{sec:rw-mcr}

Empirical work has examined MCR at the change level, characterizing factors that shape review latency and outcomes~\cite{baysal2016investigating,kononenko2016codereview,yu2015wait,zhang2022prdecision} and the role of reviewer expectations and practice~\cite{bacchelli2013expectations,rigby2013convergent,macleod2017code}. McIntosh et al.~\cite{mcintosh2016empirical} showed that low review coverage and participation correlate with post-release defects; follow-up work examined review participation across Android, Qt, and OpenStack~\cite{thongtanunam2017review} and code ownership in MCR~\cite{thongtanunam2016revisiting}. Change size is a recurring predictor~\cite{weissgerber2008small,ram2018reviewability}, and CI plays a parallel role: prior studies analyzed its effect on review duration~\cite{cassee2020silent}, the cost of repeated builds in OpenStack~\cite{maipradit2023repeated}, and bot activity in review~\cite{wessel2022quality}.

\noindent\textbf{Key difference from our work.} Prior empirical MCR studies treat each review as an independent unit. We shift the focus to chain-level coordination (review effort, conversation, and dependency-graph evolution when a change belongs to a relation chain) and identify intra-chain structural factors such as chain position, base-change activity, and structural evolution that change-level analyses cannot observe.

\subsection{Reviewer Recommendation and Review Automation}
\label{sec:rw-recommendation}

Two adjacent lines of work aim to reduce review effort through tooling. \emph{Reviewer recommendation} systems suggest reviewers based on file-location history~\cite{thongtanunam2015whoreview} or expertise and workload~\cite{mirsaeedi2020mitigating}, while pull-request decision factors have been studied on GitHub~\cite{zhang2022prdecision}. \emph{LLM-based review automation} extends this with pre-trained models that generate or refine review comments~\cite{li2022automating,tufano2022using,guo2024exploring}. Related work has built predictive models on change-level features for merged-vs-abandoned classification~\cite{chouchen2024multicr}, completion-time prediction~\cite{chouchen2023learning}, and effort-aware prioritization~\cite{chouchen2024effortaware}.

\noindent\textbf{Key difference from our work.} Existing tools operate on individual changes and do not consume chain-level signals such as chain position, base-change review state, or evolving dependency graphs. Our study shows the importance of accounting for this dependency context when reasoning about review outcomes, providing a concrete extension point for chain-aware tooling.

\subsection{Change Decomposition and Dependent Reviews}
\label{sec:bg-decomposition}

The benefits of smaller review units create a tension: focused changes improve reviewability, but real engineering tasks are often too large for a single review. Prior work has studied \emph{change decomposition}: Barnett et al.~\cite{barnett2015decomposition} proposed automated decomposition for tangled changesets at Microsoft, and Di Biase et al.~\cite{dibiase2019decomposition} showed decomposed changes reduce false-positive review comments and improve review quality. Related work on dependencies: Arabat and Sayagh~\cite{arabat2024cross} found inter-component dependencies persist in microservice systems; Arabat et al.~\cite{arabat2025ml} examined explicit \texttt{Depends-On}/\texttt{Needed-By} annotations in OpenStack; \emph{patch linkages}~\cite{hirao2019linkage,wang2021linkage,wang2022crosspatch} trace references between review discussions; and \emph{change coupling}~\cite{gall1998detection,zimmermann2005mining} mines version histories for co-evolving files.

\noindent\textbf{Key difference from our work.} Decomposition research treats decomposition as a property of individual changes; dependency research models dependencies as pairwise or textual. Relation chains are the mechanism by which developers implement decomposition in practice, introducing structural dynamics (base-change effects on descendants, merge-ordering constraints, dependency-graph evolution) that change-level studies cannot observe.

\section{Study Design}
\label{sec:methodology}

\begin{figure}[!ht]
\centering
\includegraphics[width=0.9\textwidth]{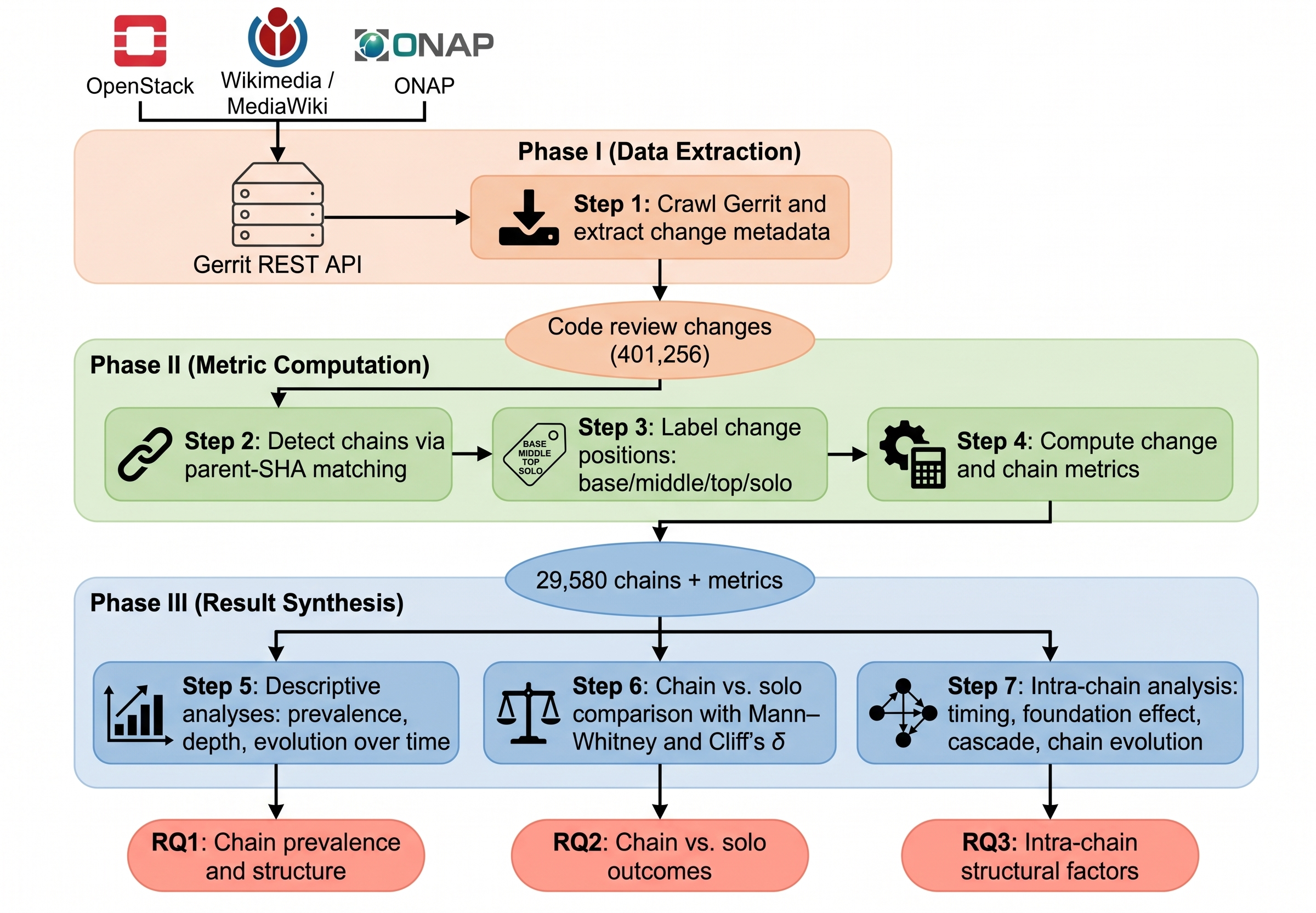}
\caption{Study methodology overview.}
\label{fig:methodology}
\end{figure}

Figure~\ref{fig:methodology} shows an illustration of our study. 

\subsection{Research Questions}
\label{sec:rqs}
We organize our study around three research questions that progressively examine relation chains as code-review artifacts: how developers adopt them across projects and over time, how chain membership relates to review outcomes compared to isolated changes, and which intra-chain structural factors shape those outcomes.
\begin{description}
\item[RQ1.] \emph{[Chain Usage]} What characterizes the usage of relation chains across projects and ecosystems?
\item[RQ2.] \emph{[Chain vs.\ Solo Outcomes]} How do relation chains reshape review coordination, validation workload, and review latency compared to isolated changes?
\item[RQ3.] \emph{[Intra-chain Structural Factors]} What structural factors within a chain shape its review outcomes?
\end{description}

\noindent
RQ1 establishes how often, how deeply, and over what time horizon developers use chains. The basic shape of chain usage (prevalence, depth, temporal stability) determines whether chain-level analysis is broadly relevant or confined to a narrow slice within a few projects. RQ2 examines whether reviews that belong to a relation chain differ systematically from isolated reviews in terms of review duration, rebasing activity, discussion, and CI workload. We compare chain and solo changes within the same size categories and inspect whether similar patterns appear across projects. If chain-associated reviews consistently differ from isolated reviews, then chain-level signals may be useful for future review analytics and tooling. Because the study is observational, we interpret the results as recurring associations rather than direct causal effects of chaining. RQ3 then examines which \emph{intra-chain} features drive the differences identified in RQ2 (chain position, base-change activity, structural evolution), turning the chain from a binary attribute into a set of fine-grained signals that downstream review-support tools (recommendation systems, dashboards, CI schedulers) could consume. Limitations are discussed in Section~\ref{sec:threats}.

\subsection{Phase~I: Data Extraction}
\label{sec:phase1}

We selected three Gerrit ecosystems (OpenStack, Wikimedia, ONAP) as representative large ecosystems where coordination across dependent changes is needed, and where prior MCR research has established the platform as a reliable empirical setting~\cite{mcintosh2016empirical,hirao2019linkage,thongtanunam2017review,baysal2016investigating}. The three differ on dimensions likely to interact with chain behaviour: review culture, language and domain, and CI-gating strictness, letting us separate effects that are general properties of chains from those tied to a specific development context. All three use the same review platform with the same chain-detection mechanism (patchset-level parent-SHA links), so cross-ecosystem comparisons rest on a uniform data model. Within each ecosystem, we required publicly accessible review history, at least 1{,}000~completed changes, and active development spanning multiple years, following prior MCR conventions~\cite{mcintosh2016empirical,hirao2019linkage,arabat2024cross}. Table~\ref{tab:corpus} summarizes the corpus. For each project, we queried the Gerrit REST API in five passes covering change metadata, patchset revisions with parent SHAs, file-level diffs, review messages (distinguishing humans from CI bots), and review labels.

\begin{table}[!ht]
\fontsize{6}{5}\selectfont
\tabcolsep=0.1cm
\caption{Overview of the studied Gerrit corpus across the OpenStack, Wikimedia, and ONAP ecosystems. Per-project counts of total changes, merged and abandoned outcomes, and reconstructed relation chains.}
\label{tab:corpus}
\centering
\begin{tabular}{llrrrr}
\hline
\textbf{Ecosystem} & \textbf{Project} & \textbf{Changes} & \textbf{Merged} & \textbf{Aband.} & \textbf{Chains} \\
\hline
OpenStack  & nova         & 42{,}781 & 31{,}145 & 11{,}636 & 3{,}682 \\
           & neutron      & 29{,}151 & 22{,}430 & 6{,}603  & 2{,}015 \\
           & cinder       & 18{,}797 & 13{,}793 & 4{,}094  & 1{,}104 \\
           & horizon      & 13{,}654 & 10{,}503 & 3{,}058  & 406     \\
           & heat         & 12{,}446 & 10{,}047 & 2{,}299  & 498     \\
\hline
Wikimedia  & mw/core      & 84{,}626 & 72{,}884 & 10{,}276 & 10{,}036 \\
           & ext/Wikibase & 28{,}839 & 25{,}692 & 2{,}972  & 3{,}662  \\
           & ext/VE       & 17{,}194 & 16{,}080 & 986      & 1{,}291  \\
           & ext/MF       & 15{,}365 & 13{,}830 & 1{,}501  & 1{,}341  \\
           & ops/puppet$^\dagger$   & 118{,}257& 110{,}366& 7{,}891  & 3{,}012 \\
\hline
ONAP       & so           & 5{,}208  & 4{,}439  & 699      & 680     \\
           & sdc          & 4{,}728  & 4{,}157  & 543      & 846     \\
           & oom          & 5{,}743  & 4{,}812  & 904      & 588     \\
           & cps          & 2{,}524  & 2{,}264  & 243      & 184     \\
           & ccsdk/cds    & 1{,}943  & 1{,}809  & 131      & 235     \\
\hline
\multicolumn{2}{l}{\textbf{Total}} & \textbf{401{,}256} & & & \textbf{29{,}580} \\
\hline
\end{tabular}

\noindent{\scriptsize $^\dagger$ops/puppet's chain count includes one auto-generated chain of depth 59{,}945 from its config-as-code workflow; we exclude it from aggregate statistics.}
\end{table}

\subsection{Phase II: Chain Detection and Metric Computation}
\label{sec:phase2}

A \emph{relation chain} is a set of Gerrit changes linked by parent--child dependencies at the \emph{patchset} level. Every patchset is a full git commit with a parent SHA; Gerrit detects a chain link whenever a patchset's parent SHA matches the latest revision SHA of another in-review change in the same project. Because each change has its own evolving patchset sequence, the parent SHA can change across patchsets, the mechanism by which chains restructure during review (Section~\ref{sec:rq3}). For chain detection (step~2), we follow Gerrit's convention and link changes by the parent SHA of each change's latest patchset. A change with no in-review parent and no in-review children is classified as \emph{solo}.

We label each chain member by position: the \emph{base} has no in-review parent, the \emph{top} has no in-review descendant, and \emph{middle} members are neither. In a chain of $n \geq 3$, exactly one base, one top, and $n-2$ middle members exist; branching chains produce one top per branch. \emph{Solo} is a fourth label for changes outside any chain. For each change we compute six \emph{per-change} metrics (step~3). \textbf{\#Revisions} is the number of distinct patchset revisions, from the Gerrit \texttt{revisions} endpoint~\cite{mcintosh2016empirical,hirao2019linkage,thongtanunam2016revisiting}. \textbf{\#Rebases} is the subset of patchsets whose \texttt{kind} is \texttt{NO\_CHANGE} or \texttt{TRIVIAL\_REBASE}, isolating dependency-driven rebases from substantive revisions. \textbf{Review duration (hours)} is the time from first patchset upload to final state, computed across merged and abandoned changes to avoid survivorship bias~\cite{baysal2016investigating,yu2015wait}. \textbf{Discussion messages} is the count of top-level human review messages, excluding accounts on a per-project bot allowlist (top-50 commenters, CI usernames such as \texttt{zuul}, \texttt{jenkins})~\cite{mcintosh2016empirical,wessel2022quality}. \textbf{Inline comments} is the count of human-authored code-anchored comments, filtered with the same allowlist. \textbf{CI/CD jobs} is the count of verification messages from accounts in the bot allowlist~\cite{maipradit2023repeated,cassee2020silent}. To account for change size as a confound, changes are bucketed by Gerrit size category (XS:~$\leq 9$ lines; S:~10--29; M:~30--99; L:~100--999; XL:~$\geq 1000$)~\cite{mcintosh2016empirical,hirao2019linkage}.

At the chain level (step~4) we compute four \emph{per-chain} metrics. \emph{Chain depth} is the number of members ($\geq 2$). \emph{Submission gap} and \emph{merge gap} are computed for each consecutive pair $(c_i, c_{i+1})$ as the time between uploads and between merges. For each chain with at least two merged members, we compute the \emph{base-change} and \emph{descendant-mean} values of each per-change metric, supporting correlation analyses. For chain-evolution analysis we compare each member's parent SHA at Patchset~1 against its parent SHA at the final patchset, classifying any non-trivial-rebase change as a structural evolution event.

\subsection{Phase III: Analysis}
\label{sec:phase3}

We use non-parametric tests throughout, chosen for robustness to the right-skewed distributions of
code-review metrics~\cite{arcuri2011practical,arcuri2014hitchhiker}. All tests are two-sided
(null hypothesis: both samples come from the same population).

For RQ1's temporal trend (step~5), monthly prevalence series violate the independence assumption
of standard correlation tests, as consecutive months share contributors, release cadences, and
backlogs. We assess monotonic trends with the Mann--Kendall test~\cite{conover1999nonparametric}
(robust to autocorrelation) and report Sen's slope~\cite{sen1968estimates} in percentage points per
year (pp/yr). Both are established in SE temporal studies~\cite{vasilescu2015quality,hassan2009predicting}.
$H_0$: no monotonic trend; $H_1$: increasing or decreasing trend.

For chain‑vs‑solo two‑sample comparisons (RQ2, step~6), we report Mann--Whitney
U~\cite{mann1947test,conover1999nonparametric} and Cliff's~$\delta$~\cite{cliff1993dominance}
with thresholds from Romano et al.~\cite{romano2006exploring}: $|\delta| < 0.147$ negligible,
$<0.33$ small, $<0.474$ medium, $\ge 0.474$ large. $H_0$: a random chain member and a random
solo change have equal probability of being larger; $H_1$: one is stochastically greater.
Given the large sample sizes, significance alone can be misleading; we therefore interpret
Cliff's $\delta$ as the primary estimate of practical magnitude and focus on directional
consistency across projects and ecosystems.

For per‑chain base‑vs‑descendant correlations (RQ3, step~7), we report Spearman's~$\rho$
with two‑sided $p$‑values~\cite{conover1999nonparametric}, which captures monotonic association
without assuming linearity or normality. $H_0$: base‑change and descendant‑mean metrics are
independent in rank; $H_1$: they are monotonically associated.

Significance markers: \textsuperscript{***}~$p<0.001$, \textsuperscript{**}~$p<0.01$,
\textsuperscript{*}~$p<0.05$, ns otherwise. No multiple‑comparison correction is applied;
each project is an independent observational unit, and we interpret the sign and magnitude
of each effect individually.

\section{Results}
\label{sec:results}

\subsection{RQ1: Prevalence and Structure}
\label{sec:rq1}

\begin{figure}[!ht]
\centering
\includegraphics[width=0.9\textwidth,height=3cm]{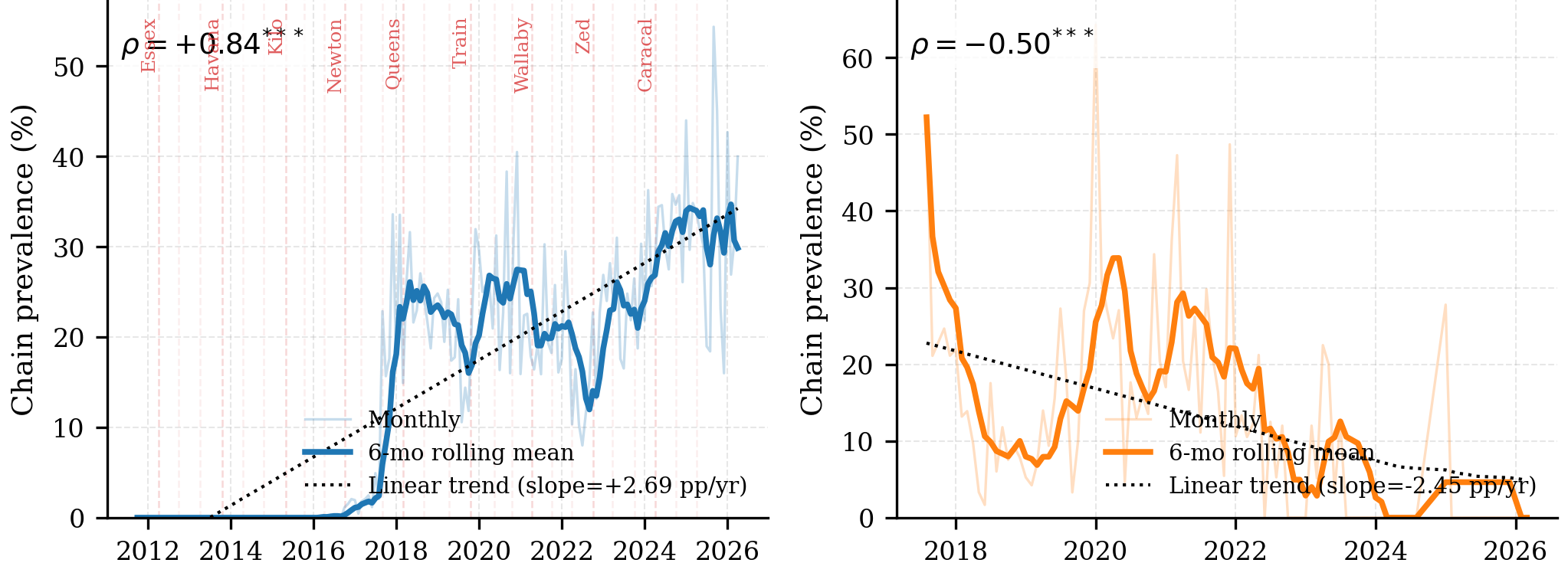}
\caption{Monthly chain-prevalence exemplars: (a)~increasing trend in OpenStack Neutron; (b)~decreasing trend in ONAP \texttt{oom}. Bold line: 6-month rolling mean; dotted line: linear trend.}
\label{fig:prevalence}
\end{figure}

\begin{figure}[!ht]
\centering
\includegraphics[width=0.9\textwidth,height=4.7cm]{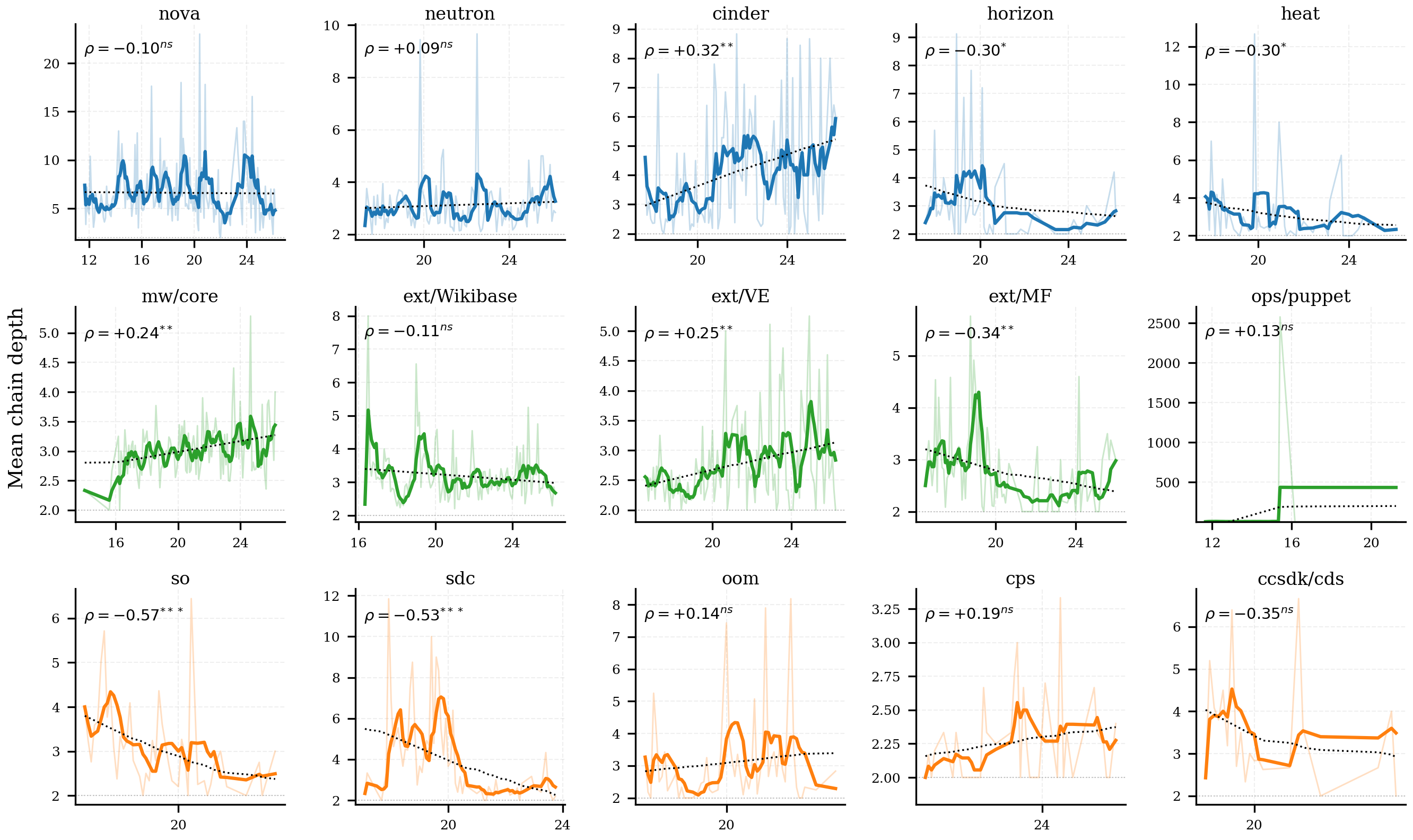}
\caption{Monthly mean chain depth per project (6-month rolling mean; dotted line = linear trend). Horizontal dotted line at depth~2 marks the pair floor.}
\label{fig:depth-evolution}
\vspace{-0.6cm}
\end{figure}

Table~\ref{tab:prevalence-trends} summarizes per-project chain-prevalence trends; Figure~\ref{fig:prevalence} illustrates two contrasting exemplars. Table~\ref{tab:depth-stats} reports per-project chain-depth statistics, and Figure~\ref{fig:depth-evolution} shows monthly mean chain depth per project.

\begin{table}[!ht]
\centering
\fontsize{6}{5}\selectfont
\tabcolsep=0.1cm
\caption{Per-project chain-prevalence trends (Mann--Kendall test on monthly prevalence series). \emph{Chained \%} is the proportion of project changes that belong to a relation chain. Sen's slope reports the trend magnitude in percentage points per year.}
\label{tab:prevalence-trends}
\begin{tabular}{l l r r r l}
\hline
Ecosystem & Project & Chained \% & Months & Sen's slope (pp/yr) & MK $p$ \\
\hline
OpenStack & Nova         & 31.5\% & 175 & $+1.83$ & \textsuperscript{***} \\
          & Neutron      & 10.8\% & 176 & $+2.44$ & \textsuperscript{***} \\
          & Cinder       & 11.0\% & 168 & $+2.81$ & \textsuperscript{***} \\
          & Horizon      &  5.1\% & 174 & $+0.76$ & \textsuperscript{***} \\
          & Heat         &  7.5\% & 159 & $+2.23$ & \textsuperscript{***} \\
\hline
Wikimedia & mw/core            & 19.1\% & 170 & $+1.93$ & \textsuperscript{***} \\
          & ext/Wikibase       & 21.9\% & 167 & $+2.76$ & \textsuperscript{***} \\
          & ext/VisualEditor   & 13.2\% & 170 & $+1.54$ & \textsuperscript{***} \\
          & ext/MobileFrontend & 12.1\% & 170 & $+0.43$ & \textsuperscript{***} \\
          & ops/puppet$^\dagger$ & 97.3\% & 170 & $+0.27$ & \textsuperscript{***} \\
\hline
ONAP      & so        & 27.1\% & 81  & $-3.42$ & \textsuperscript{**}  \\
          & sdc       & 48.5\% & 89  & $-8.02$ & \textsuperscript{***} \\
          & oom       & 16.9\% & 104 & $-1.99$ & \textsuperscript{***} \\
          & cps       & 11.2\% & 65  & $-1.53$ & \textsuperscript{*}   \\
          & ccsdk/cds & 25.0\% & 52  & $-1.59$ & \ \,ns                \\
\hline
\end{tabular}
\smallskip

\noindent{\scriptsize $^\dagger$\texttt{ops/puppet}'s near-100\% prevalence reflects auto-generated config-as-code chains, not human-authored review.}
\end{table}
\vspace{-0.3cm}

\textbf{Finding \#1.1: Relation-chain prevalence varies across projects and ecosystems.}
Excluding the \texttt{operations/puppet} outlier, the proportion of changes in a chain ranges from 5.1\% in OpenStack \texttt{Horizon} to 48.5\% in ONAP \texttt{sdc}, with a median of 17\% across the remaining 14 projects. Variation differs by ecosystem: ONAP 11--49\%, OpenStack 5--32\%, Wikimedia 12--22\%. Chain usage is therefore not uniform but reflects differences in development practices and tooling conventions.

\textbf{Finding \#1.2: Relation-chain adoption evolves over time, with most projects showing increasing usage.}
The Mann--Kendall trend test (Table~\ref{tab:prevalence-trends}) yields significant monotonic trends in 14 of 15 projects: 10 increasing, 4 decreasing. The only project without a significant trend is \texttt{ccsdk/cds}, which has the shortest observation window (38 months). All OpenStack projects exhibit increasing trends ($+0.8$ to $+2.8$~pp/yr, $p<0.001$), and four of five Wikimedia projects follow the same pattern. Several ONAP projects display decreasing trends ($-8.0$ to $-1.5$~pp/yr), notably \texttt{oom} (Figure~\ref{fig:prevalence}b), where monthly prevalence drops from over 60\% in 2018--2019 to below 20\% by 2024.
\vspace{-0.3cm}

ONAP \texttt{oom} change~137529,\footnote{\url{https://gerrit.onap.org/r/c/oom/+/137529}} which introduces Kafka support across multiple policy charts in a single 2024 change, illustrates this evolution. Similar cross-component modifications in earlier years were often decomposed into linked chains, suggesting a gradual shift away from chain-oriented coordination; we revisit potential explanations in Section~\ref{sec:discussion}.

\textbf{Finding \#1.3: Relation chains are dominated by short pair-like structures, with project-specific deep tails and one auto-generated outlier.}
Per-project chain-depth statistics show similar central tendencies across ecosystems (Table~\ref{tab:depth-stats}): 12 of 15 projects have median chain depth 2, the first quartile is 2 in 14 of 15, and the third quartile is at most 4 in 11 of 15. Relation-chain activity therefore consists mostly of pair-like structures: a single dependent change on an in-review parent, or sequences of two to four members. Mean per-chain depth is correspondingly low (2.3--6.1), and standard deviation in 12 of 15 projects is below 4. The tails differ. Maximum chain depth ranges from 5 (\texttt{cps}) to 98 (\texttt{sdc}), with seven projects supporting at least one chain of depth $>$30. ONAP \texttt{sdc} has the deepest profile (mean 6.1, Q3 of 7, longest 98 members); at the other extreme, \texttt{cps} has maximum depth 5 and standard deviation 0.6. The \texttt{ops/puppet} project is an outlier: its maximum of 59{,}945 reflects a single auto-generated chain from its config-as-code workflow (Table~\ref{tab:corpus}), and we exclude it from aggregates.

\begin{table}[H]
\centering
\fontsize{6}{5}\selectfont
\tabcolsep=0.1cm
\caption{Per-project chain-depth statistics. \texttt{ops/puppet}'s maximum reflects a single auto-generated config-as-code chain (Table~\ref{tab:corpus}).}
\label{tab:depth-stats}
\begin{tabular}{llcccccccc}
\toprule
Ecosystem & Project & $n$ & Min & Q1 & Median & Mean & Q3 & Max & Std \\
\midrule
OpenStack & nova & 4,985 & 2 & 2 & 2 & 2.7 & 3 & 56 & 1.8 \\
 & neutron & 2,015 & 2 & 2 & 2 & 3.1 & 3 & 26 & 2.5 \\
 & cinder & 1,104 & 2 & 2 & 2 & 4.8 & 5 & 38 & 5.4 \\
 & horizon & 406 & 2 & 2 & 2 & 3.6 & 4 & 22 & 3.3 \\
 & heat & 498 & 2 & 2 & 2 & 3.5 & 4 & 29 & 3.0 \\
\midrule
Wikimedia & mw/core & 10,036 & 2 & 2 & 2 & 3.2 & 4 & 31 & 2.1 \\
 & ext/Wikibase & 3,662 & 2 & 2 & 2 & 3.4 & 4 & 34 & 2.8 \\
 & ext/VisualEditor & 1,291 & 2 & 2 & 2 & 3.0 & 3 & 12 & 1.6 \\
 & ext/MobileFrontend & 1,341 & 2 & 2 & 3 & 3.4 & 4 & 22 & 2.3 \\
 & ops/puppet & 3,012 & 2 & 3 & 5 & 38.2 & 9 & 59,945 & 1244.4 \\
\midrule
ONAP & so & 680 & 2 & 2 & 2 & 3.2 & 4 & 21 & 2.3 \\
 & sdc & 846 & 2 & 2 & 3 & 6.1 & 7 & 98 & 7.8 \\
 & oom & 588 & 2 & 2 & 2 & 3.4 & 3 & 46 & 3.6 \\
 & cps & 184 & 2 & 2 & 2 & 2.3 & 2 & 5 & 0.6 \\
 & ccsdk/cds & 235 & 2 & 2 & 2 & 3.5 & 4 & 18 & 2.7 \\
\bottomrule
\end{tabular}
\end{table}

\textbf{Finding \#1.4: Mean chain depth shows no systematic temporal trend in most projects.}
Grouping each chain by the month its base was created and computing monthly mean depth (Figure~\ref{fig:depth-evolution}), Spearman correlations between calendar month and mean depth fall in $-0.34$ to $+0.33$ and are non-significant in 12 of 15 projects. The exceptions are two ONAP projects, \texttt{so} ($\rho=-0.57$, $p<0.001$) and \texttt{sdc} ($\rho=-0.53$, $p<0.001$), where chains have become shallower over time, consistent with their declining prevalence in Finding~\#1.2. OpenStack Cinder shows the opposite pattern ($\rho=+0.32$, $p<0.01$). Outside these three cases, chain \emph{shape} is stable while \emph{adoption} grows: the increasing prevalence in Finding~\#1.2 reflects more chains, not longer ones.

\textbf{Finding \#1.5: Depth spikes correspond to coordinated engineering campaigns, not noise.}
The temporal trends mask a recurring pattern in Figure~\ref{fig:depth-evolution}: individual months in which mean chain depth jumps to several times the project's long-run median. We identified 14 such spike months across the corpus (mean depth above the project's median plus two standard deviations); 13 are multi-chain bursts where three to nineteen chains, several deep, are created together. One example is OpenStack Horizon in December~2018, when seventeen chains were created in one month with a mean depth of nine and a maximum of twenty-two: a single author decomposing a project-wide \texttt{pylint} cleanup\footnote{\url{https://review.opendev.org/q/topic:pylint+project:openstack/horizon}} into many dependent chains. Similar patterns recur for architectural migrations and refactoring sweeps (full list in the replication package): depth spikes mark coordinated multi-chain campaigns.
\vspace{-0.3cm}

\begin{rqsummary}{1}
Relation chains are a common but ecosystem coordination mechanism in Gerrit-based development, with prevalence ranging from 5\% to 49\% across projects (median 17\%). Chain adoption is increasing in most projects. Chains are typically shallow (median depth 2), and their structural characteristics remain stable despite changing adoption rates. Deeper chains, when they occur, are associated with engineering workflows such as sequential refactoring and diagnostic branching.
\end{rqsummary}
\vspace{-0.3cm}
\subsection{RQ2: Review Outcomes}
\label{sec:rq2}
\begin{figure}[ht!]
\centering
\includegraphics[width=\textwidth,height=6cm]{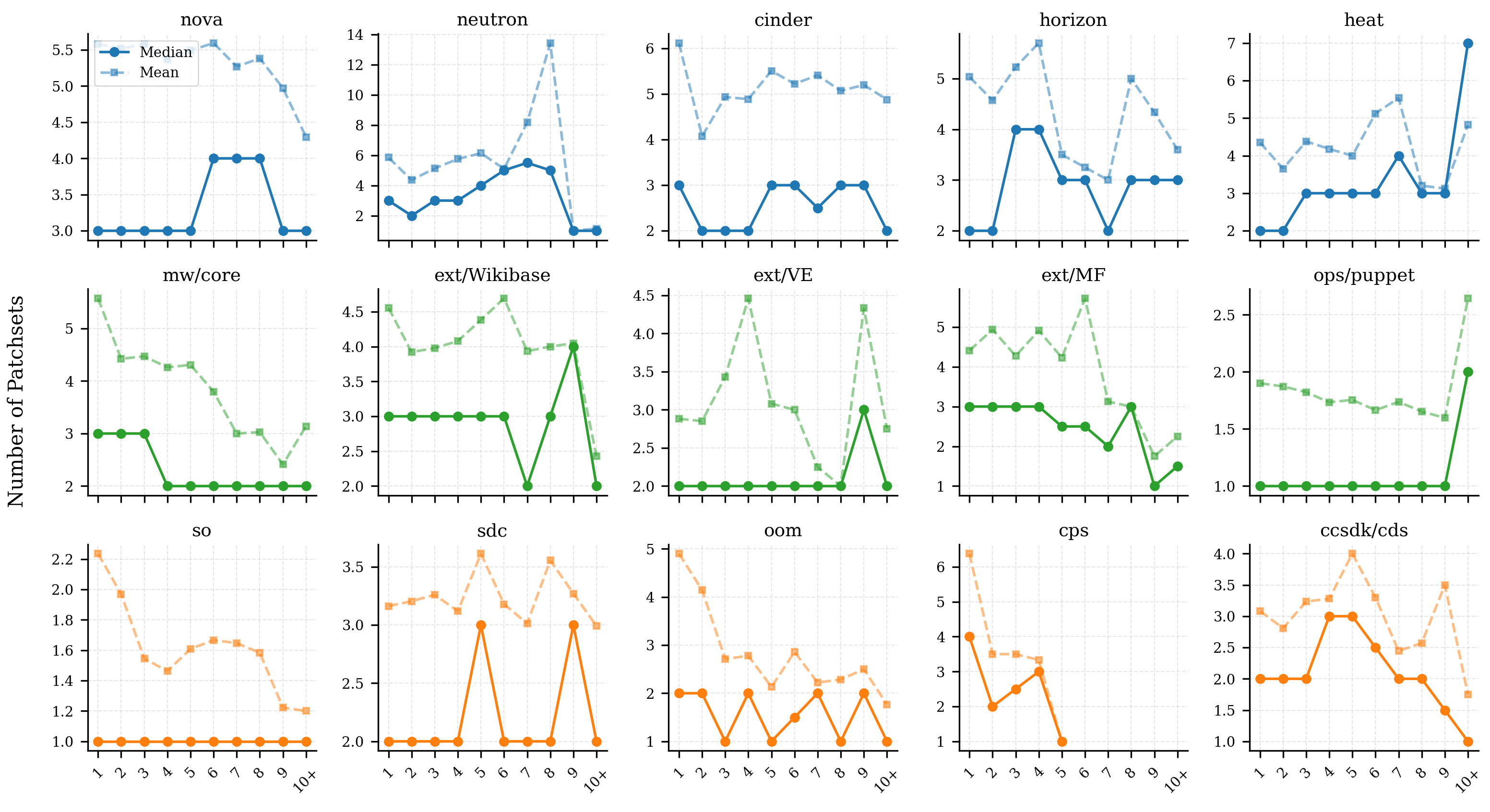}
\caption{Patchset count by chain position-from-base, per project. Solid line: median; dashed line: mean.}
\label{fig:position-patchsets}
\end{figure}

\vspace{-0.2cm}
Table~\ref{tab:rq2-ecosystems-full} reports descriptive statistics and Mann--Whitney effect sizes for chain-vs-solo comparisons across six metrics and five Gerrit size buckets. Table~\ref{tab:rq2-position-full} breaks down the same metrics by chain-position label . Figure~\ref{fig:position-patchsets} shows per-project patchset count by chain position.

\textbf{Finding \#2.1: Relation chains introduce synchronization overhead through additional revisions, rebasing activity, and longer review duration.} Across all five Gerrit size buckets, chain members take a median of $2.6\times$ longer to merge and undergo $1.5\times$ more revision rounds than solo changes of comparable size (Mann--Whitney U, $p<0.001$; Table~\ref{tab:rq2-ecosystems-full}). Cliff's~$\delta$ is \emph{small} in four of five buckets and reaches \emph{large} only in the XL bucket ($\delta = 0.49$), indicating that the overhead is concentrated among very large changes rather than reflecting a full separation of distributions. This additional cost reflects synchronization overhead in dependency-linked review workflows, where developers must repeatedly coordinate rebasing, validation, and merge ordering across dependent changes. The OpenStack Horizon December~2018 \texttt{pylint} cleanup\footnote{\url{https://review.opendev.org/q/topic:pylint+project:openstack/horizon}}, which created seventeen dependent chains in a single month, illustrates this effect: each chain accumulated not only its own review iterations, but also coordination overhead inherited from its dependent review structure.

\textbf{Finding \#2.2: Chain members do not attract more top-level discussion than solo changes, but receive significantly more code-anchored inline comments in OpenStack and Wikimedia.}
Top-level discussion volume shows no chain-vs-solo pattern: in OpenStack and Wikimedia, solo changes attract more median discussion messages than middle- or top-of-chain members (Table~\ref{tab:rq2-position-full}). Inline comments tell a different story. In OpenStack, chain members receive significantly more inline comments than size-matched solo changes across all five size buckets (Cliff's $\delta = 0.21$--$0.34$, all \emph{small}, $p<0.001$; Table~\ref{tab:rq2-ecosystems-full}). Wikimedia shows significant differences in every bucket but with mixed direction: chain members receive more inline comments in some buckets and fewer in others, an ecosystem-specific interaction rather than a uniform gap. ONAP shows the smallest and least consistent effect. The picture is consistent with reviewers shifting the form of feedback: top-level discussion is split across chain members, while code-anchored scrutiny stays on the changes containing the dependent edits, particularly under strict gating. The observed pattern is consistent with reviewers allocating more activity toward synchronization and validation maintenance rather than additional top-level conversational discussion.

\textbf{Finding \#2.3: Relation chains exhibit a CI amplification effect under strict gating regimes.}
Chain members run a median of $10$--$23$ CI/CD jobs depending on position, against fewer than two for solo changes in the same ecosystem (Table~\ref{tab:rq2-position-full}). The split is largest in OpenStack, where base, middle, and top members each trigger a test pipeline on every patchset under the project's gating policy~\cite{maipradit2023repeated,cassee2020silent}; the accumulation matches the multi-hour merge gaps reported in Finding~\#3.1. ONAP, with lighter CI gating, shows the smallest split (chain CI medians of $0$--$2$ versus solo medians of $2$). The OpenStack Neutron \texttt{[OVN]} migration of November~2019\footnote{\url{https://review.opendev.org/q/topic:ovn-migration+project:openstack/neutron}}, with nine dependent chains advancing the same migration, illustrates the cumulative cost: each member's revisions retriggered the CI suite. We refer to this recurring increase in downstream validation workload as the \emph{CI amplification effect}, where revisions to one chain member trigger repeated validation across dependent descendants.

\begin{table}[H]
\centering

\caption{Per-ecosystem chain-position breakdown of six review metrics.
For each ecosystem (OpenStack, Wikimedia, ONAP), each metric is split across
the three chain-position labels (Base, Middle, Top) and summarised as
Min, Q1, Median, Mean, Q3, Max. \texttt{ops/puppet} is excluded from the
Wikimedia rows (see Table~\ref{tab:corpus}).}
\label{tab:rq2-position-full}

\scriptsize
\renewcommand{\arraystretch}{1.1}

\resizebox{\textwidth}{!}{%

\begin{tabular}{c l *{18}{r}}

\toprule

& & \multicolumn{6}{c}{\textbf{Base}}
& \multicolumn{6}{c}{\textbf{Middle}}
& \multicolumn{6}{c}{\textbf{Top}} \\

\cmidrule(lr){3-8}
\cmidrule(lr){9-14}
\cmidrule(lr){15-20}

& Metric
& Min & Q1 & Med & Mean & Q3 & Max
& Min & Q1 & Med & Mean & Q3 & Max
& Min & Q1 & Med & Mean & Q3 & Max \\

\midrule

\multirow{6}{*}{\rotatebox[origin=c]{90}{\textbf{OpenStack}}}

& \#Revisions
& 1 & 1 & 3 & 5.61 & 6 & 126
& 1 & 2 & 3 & 5.72 & 7 & 103
& 1 & 1 & 3 & 4.78 & 5 & 114 \\

& \#Rebases
& 0 & 0 & 2 & 4.61 & 5 & 125
& 0 & 1 & 2 & 4.72 & 6 & 102
& 0 & 0 & 2 & 3.78 & 4 & 113 \\

& Review dur.\,(h)
& 0.18 & 63.9 & 270 & 1167 & 990 & 60,844
& 0.28 & 116 & 362 & 1181 & 1143 & 47,431
& 0.40 & 105 & 324 & 1011 & 964 & 38,658 \\

& Discussion msg.
& 0 & 1 & 8 & 30.4 & 27 & 1,796
& 0 & 0 & 4 & 25.4 & 19 & 951
& 0 & 0 & 4 & 15.6 & 16 & 1,185 \\

& Inline comments
& 0 & 7 & 17 & 38.6 & 40 & 1,786
& 0 & 9 & 19 & 36.1 & 39 & 935
& 0 & 7 & 15 & 29.1 & 33 & 1,185 \\

& CI/CD jobs
& 0 & 5 & 10 & 21.5 & 23.2 & 550
& 0 & 6 & 14 & 23.0 & 28 & 408
& 0 & 5 & 11 & 19.2 & 23 & 581 \\

\midrule

\multirow{6}{*}{\rotatebox[origin=c]{90}{\textbf{Wikimedia}}}

& \#Revisions
& 1 & 1 & 3 & 5.04 & 6 & 180
& 1 & 2 & 3 & 4.85 & 6 & 121
& 1 & 1 & 2 & 3.56 & 4 & 198 \\

& \#Rebases
& 0 & 0 & 2 & 4.04 & 5 & 179
& 0 & 1 & 2 & 3.85 & 5 & 120
& 0 & 0 & 1 & 2.56 & 3 & 197 \\

& Review dur.\,(h)
& 0 & 2.95 & 33.5 & 489 & 176 & 85,686
& 0.00 & 4.75 & 43.9 & 333 & 155 & 43,391
& 0 & 4.56 & 31.8 & 366 & 147 & 47,118 \\

& Discussion msg.
& 2 & 10 & 16 & 26.1 & 29 & 765
& 2 & 10 & 16 & 23.4 & 27 & 505
& 1 & 9 & 13 & 17.8 & 21 & 740 \\

& Inline comments
& 0 & 2 & 6 & 10.1 & 11 & 405
& 0 & 2 & 5 & 9.61 & 12 & 312
& 0 & 1 & 4 & 7.50 & 10 & 380 \\

& CI/CD jobs
& 0 & 3 & 9 & 16.0 & 18 & 547
& 0 & 3 & 9 & 13.8 & 17 & 353
& 0 & 3 & 7 & 10.3 & 13 & 530 \\

\midrule

\multirow{6}{*}{\rotatebox[origin=c]{90}{\textbf{ONAP}}}

& \#Revisions
& 1 & 1 & 2 & 3.59 & 4 & 77
& 1 & 2 & 2 & 2.95 & 3 & 50
& 1 & 1 & 2 & 2.78 & 3 & 41 \\

& \#Rebases
& 0 & 0 & 1 & 2.59 & 3 & 76
& 0 & 1 & 1 & 1.95 & 2 & 49
& 0 & 0 & 1 & 1.78 & 2 & 40 \\

& Review dur.\,(h)
& 0.00 & 2.85 & 24.4 & 203 & 146 & 6,353
& 0.01 & 3.52 & 29.6 & 120 & 117 & 4,625
& 0.02 & 3.66 & 23.3 & 122 & 116 & 2,562 \\

& Discussion msg.
& 2 & 8 & 13 & 30.3 & 29 & 820
& 2 & 8 & 11 & 16.0 & 16 & 567
& 2 & 7 & 11 & 20.4 & 19 & 655 \\

& Inline comments
& 0 & 7 & 11 & 25.0 & 23 & 748
& 1 & 8 & 9 & 14.1 & 14 & 511
& 0 & 6 & 9 & 17.2 & 16 & 604 \\

& CI/CD jobs
& 0 & 0 & 2 & 5.22 & 6 & 347
& 0 & 0 & 0 & 1.92 & 3 & 96
& 0 & 0 & 2 & 3.16 & 4 & 107 \\

\bottomrule

\end{tabular}

} 

\end{table}

%

\begin{table}[H]
\scriptsize
\tabcolsep=0.02cm
\centering
\caption{Per‑ecosystem descriptive statistics and effect sizes for review metrics,
         comparing chain (Ch) vs.\ solo (So) changes across Gerrit size buckets.
         $p$‑values from Mann–Whitney U test;
         Cliff's $\delta$ magnitude: N\,=\,negligible, S\,=\,small, M\,=\,medium, L\,=\,large.}
\label{tab:rq2-ecosystems-full}
\resizebox{0.95\textwidth}{!}{%
\setlength{\tabcolsep}{1pt}
\renewcommand{\arraystretch}{1.05}

\begin{tabular}{c l c rr rr rr rr rr rr r r}
\toprule
& & &
\multicolumn{2}{c}{Min} &
\multicolumn{2}{c}{Q1} &
\multicolumn{2}{c}{Mean} &
\multicolumn{2}{c}{Median} &
\multicolumn{2}{c}{Q3} &
\multicolumn{2}{c}{Max} \\
\cmidrule(lr){4-5}\cmidrule(lr){6-7}\cmidrule(lr){8-9}
\cmidrule(lr){10-11}\cmidrule(lr){12-13}\cmidrule(lr){14-15}
Ecosystem & Metric & Size &
Ch & So & Ch & So & Ch & So & Ch & So & Ch & So & Ch & So &
$p$ & Cliff's $\delta$ \\
\midrule

\multirow{30}{*}[-60pt]{\rotatebox[origin=c]{90}{\textbf{OpenStack}}}
 & \multirow{5}{*}{\#Revisions}
   & XS & 1 & 1 & 1 & 1 & 2.349 & 1.889 & 1 & 1 & 3 & 2 &  56 & 229 & $<$0.001 & 0.13 (\cN) \\
 & & S  & 1 & 1 & 1 & 1 & 3.662 & 2.753 & 2 & 2 & 4 & 3 &  63 &  89 & $<$0.001 & 0.17 (\cS) \\
 & & M  & 1 & 1 & 1 & 1 & 4.772 & 4.011 & 3 & 2 & 6 & 5 &  91 & 106 & $<$0.001 & 0.10 (\cN) \\
 & & L  & 1 & 1 & 2 & 2 & 8.200 & 7.481 & 5 & 4 & 10 & 9 & 126 & 155 & $<$0.001 & 0.08 (\cN) \\
 & & XL & 1 & 1 & 3 & 1 & 10.092 & 8.222 & 6 & 3 & 12 & 10 & 105 & 176 & $<$0.001 & 0.21 (\cS) \\
\cmidrule{2-17}
 & \multirow{5}{*}{\#Rebases}
   & XS & 0 & 0 & 0 & 0 & 0.736 & 0.266 & 0 & 0 & 1 & 0 &  45 & 175 & $<$0.001 & 0.20 (\cS) \\
 & & S  & 0 & 0 & 0 & 0 & 1.318 & 0.428 & 0 & 0 & 1 & 0 &  51 &  67 & $<$0.001 & 0.27 (\cS) \\
 & & M  & 0 & 0 & 0 & 0 & 1.670 & 0.698 & 0 & 0 & 2 & 0 &  59 &  75 & $<$0.001 & 0.26 (\cS) \\
 & & L  & 0 & 0 & 0 & 0 & 2.606 & 1.432 & 1 & 0 & 3 & 1 &  65 &  66 & $<$0.001 & 0.23 (\cS) \\
 & & XL & 0 & 0 & 0 & 0 & 2.006 & 0.901 & 1 & 0 & 2.250 & 0 &  36 &  54 & $<$0.001 & 0.33 (\cS) \\
\cmidrule{2-17}
 & \multirow{5}{*}{\makecell{Review dur.\\(hours)}}
   & XS &  0.2 &  0 & 24.1 & 13.4 &  486.179 &  334.667 & 106.167 &  60.036 &  365.022 & 219.980 & 23673 & 56582.8 & $<$0.001 & 0.15 (\cS) \\
 & & S  &  0.2 &  0 & 67 & 26.2 &  798.442 &  539.740 & 213.154 & 116.154 &  672.892 & 401.990 & 32944.7 & 62328.2 & $<$0.001 & 0.20 (\cS) \\
 & & M  &  0.2 &  0 & 99.4 & 45.4 & 1067.866 &  737.522 & 323.892 & 180.072 &  982.629 & 654.940 & 41921.9 & 59373.7 & $<$0.001 & 0.18 (\cS) \\
 & & L  &  0.2 &  0 & 180.2 & 69.2 & 1646.194 & 1077.740 & 566.216 & 318.616 & 1675.422 & 1069.743 & 60844.4 & 58980.3 & $<$0.001 & 0.21 (\cS) \\
 & & XL &  0.3 &  0 & 202.3 & 10.7 & 2081.738 &  643.074 & 755.262 &  95.516 & 2108.616 &  524.222 & 47430.6 & 23719.5 & $<$0.001 & 0.48 (\cL) \\
\cmidrule{2-17}
 & \multirow{5}{*}{\makecell{Discussion\\messages}}
   & XS & 0 & 0 &  5 &  6 & 15.485 & 19.556 & 10 & 11 & 19 & 23 &  283 & 3966 & $<$0.001 & $-$0.12 (\cN) \\
 & & S  & 0 & 0 &  7 &  8 & 23.843 & 27.665 & 14 & 16 & 29 & 32 &  368 & 1777 & $<$0.001 & $-$0.05 (\cN) \\
 & & M  & 0 & 0 &  8 &  9 & 30.613 & 39.130 & 17 & 20 & 36 & 44 &  748 & 3372 & $<$0.001 & $-$0.10 (\cN) \\
 & & L  & 0 & 0 & 11 & 11 & 51.410 & 70.236 & 25 & 29 & 59 & 75 & 1348 & 3314 & $<$0.001 & $-$0.07 (\cN) \\
 & & XL & 0 & 0 & 13 &  8 & 86.777 & 84.590 & 33 & 24 & 84 & 79 & 1786 & 3525 & $<$0.001 & 0.09 (\cN) \\
\cmidrule{2-17}
 & \multirow{5}{*}{\makecell{Inline\\comments}}
   & XS & 0 & 0 & 0 &  2 & 12.529 & 17.906 &  4 & 10 & 15 & 22 &  288 & 3966 & $<$0.001 & $-$0.22 (\cS) \\
 & & S  & 0 & 0 & 0 &  4 & 14.514 & 24.324 &  3 & 12 & 15 & 29 &  580 & 1777 & $<$0.001 & $-$0.31 (\cS) \\
 & & M  & 0 & 0 & 0 &  5 & 17.201 & 33.623 &  4 & 15 & 16 & 37 &  752 & 3372 & $<$0.001 & $-$0.34 (\cM) \\
 & & L  & 0 & 0 & 2 &  8 & 37.134 & 62.364 & 10 & 23 & 34 & 62 & 1360 & 3314 & $<$0.001 & $-$0.25 (\cS) \\
 & & XL & 0 & 0 & 2 &  7 & 89.455 & 82.035 & 20 & 21 & 98.500 & 76.750 & 1796 & 3525 &    0.230 & $-$0.03 (\cN) \\
\cmidrule{2-17}
 & \multirow{5}{*}{CI/CD jobs}
   & XS & 0 & 0 & 3 & 0 & 10.463 &  4.295 &  7 & 0 & 13 &  6 &  281 &  848 & $<$0.001 & 0.44 (\cM) \\
 & & S  & 0 & 0 & 5 & 0 & 15.741 &  6.063 & 10 & 0 & 20 &  8 &  332 &  276 & $<$0.001 & 0.50 (\cL) \\
 & & M  & 0 & 0 & 6 & 0 & 19.870 &  8.637 & 12 & 0 & 24 & 10 &  387 & 1344 & $<$0.001 & 0.49 (\cL) \\
 & & L  & 0 & 0 & 6 & 0 & 30.304 & 12.266 & 16 & 0 & 37 & 10 &  581 &  592 & $<$0.001 & 0.52 (\cL) \\
 & & XL & 0 & 0 & 5 & 0 & 33.041 &  7.712 & 14 & 0 & 38.250 &  2 &  444 &  621 & $<$0.001 & 0.64 (\cL) \\

\midrule

\multirow{30}{*}[-60pt]{\rotatebox[origin=c]{90}{\textbf{Wikimedia}}}
 & \multirow{5}{*}{\#Revisions}
   & XS & 1 & 1 & 1 & 1 & 1.977 & 1.570 & 2 & 1 & 2 & 2 &  59 &  56 & $<$0.001 & 0.25 (\cS) \\
 & & S  & 1 & 1 & 1 & 1 & 2.797 & 2.188 & 2 & 1 & 3 & 3 &  53 &  81 & $<$0.001 & 0.25 (\cS) \\
 & & M  & 1 & 1 & 2 & 1 & 3.972 & 2.861 & 3 & 2 & 5 & 4 & 102 &  76 & $<$0.001 & 0.27 (\cS) \\
 & & L  & 1 & 1 & 2 & 1 & 6.587 & 3.983 & 4 & 2 & 8 & 5 & 198 & 152 & $<$0.001 & 0.35 (\cM) \\
 & & XL & 1 & 1 & 2 & 1 & 9.220 & 3.341 & 4 & 1 & 10 & 2 & 180 & 229 & $<$0.001 & 0.55 (\cL) \\
\cmidrule{2-17}
 & \multirow{5}{*}{\#Rebases}
   & XS & 0 & 0 & 0 & 0 & 0.632 & 0.161 & 0 & 0 & 1 & 0 &  36 &  29 & $<$0.001 & 0.37 (\cM) \\
 & & S  & 0 & 0 & 0 & 0 & 0.833 & 0.274 & 1 & 0 & 1 & 0 &  34 &  36 & $<$0.001 & 0.40 (\cM) \\
 & & M  & 0 & 0 & 0 & 0 & 1.031 & 0.371 & 1 & 0 & 1 & 0 &  40 &  57 & $<$0.001 & 0.38 (\cM) \\
 & & L  & 0 & 0 & 0 & 0 & 1.343 & 0.484 & 1 & 0 & 2 & 0 &  45 &  71 & $<$0.001 & 0.39 (\cM) \\
 & & XL & 0 & 0 & 0 & 0 & 1.388 & 0.338 & 1 & 0 & 2 & 0 &  30 &  50 & $<$0.001 & 0.44 (\cM) \\
\cmidrule{2-17}
 & \multirow{5}{*}{\makecell{Review dur.\\(hours)}}
   & XS & 0 & 0 & 0 & 0.2 &  89.256 & 139.776 &  0.193 &  1.576 &  13.016 &  20.944 & 47162.1 & 79040.4 & $<$0.001 & $-$0.25 (\cS) \\
 & & S  & 0 & 0 & 0 & 0.4 & 164.258 & 235.849 &  2.222 &  5.796 &  50.216 &  67.619 & 53249.6 & 52312.5 & $<$0.001 & $-$0.11 (\cN) \\
 & & M  & 0 & 0 & 0.6 & 0.3 & 291.977 & 299.865 & 19.830 & 11.239 & 119.545 &  99.448 & 59875.1 & 93770.5 & $<$0.001 & 0.07 (\cN) \\
 & & L  & 0 & 0 & 4.5 & 0.1 & 452.677 & 294.488 & 66.417 &  4.396 & 241.956 & 110.864 & 85686.3 & 77280.7 & $<$0.001 & 0.29 (\cS) \\
 & & XL & 0 & 0 & 2.6 & 0.2 & 693.852 & 145.923 & 59.204 &  0.386 & 311.383 &   6.074 & 40490.2 & 29884.2 & $<$0.001 & 0.51 (\cL) \\
\cmidrule{2-17}
 & \multirow{5}{*}{\makecell{Discussion\\messages}}
   & XS & 0 & 0 & 2 & 2 &  4.194 &  5.868 & 3 & 6 &  5 &  8 & 185 & 252 & $<$0.001 & $-$0.25 (\cS) \\
 & & S  & 0 & 0 & 2 & 2 &  5.686 &  7.119 & 4 & 6 &  7 &  9 & 143 & 231 & $<$0.001 & $-$0.12 (\cN) \\
 & & M  & 0 & 0 & 2 & 2 &  7.576 &  8.454 & 5 & 6 & 10 & 11 & 159 & 171 & $<$0.001 & $-$0.03 (\cN) \\
 & & L  & 0 & 0 & 3 & 2 & 11.941 & 10.535 & 7 & 5 & 14 & 12 & 274 & 395 & $<$0.001 & 0.12 (\cN) \\
 & & XL & 0 & 0 & 3 & 0 & 17.220 &  7.265 & 7 & 1 & 17 &  5.500 & 405 & 681 & $<$0.001 & 0.48 (\cL) \\
\cmidrule{2-17}
 & \multirow{5}{*}{\makecell{Inline\\comments}}
   & XS & 1 & 1 & 4 & 6 &  7.109 &  8.834 &  6 &  7 &  8 & 10 & 296 & 252 & $<$0.001 & $-$0.27 (\cS) \\
 & & S  & 1 & 1 & 5 & 6 & 10.309 & 11.319 &  8 &  9 & 12 & 14 & 200 & 317 & $<$0.001 & $-$0.10 (\cN) \\
 & & M  & 1 & 1 & 7 & 6 & 15.200 & 13.778 & 11 & 10 & 18 & 17 & 400 & 303 & $<$0.001 & 0.08 (\cN) \\
 & & L  & 1 & 1 & 9 & 6 & 26.174 & 17.419 & 16 &  9 & 32 & 21 & 740 & 443 & $<$0.001 & 0.28 (\cS) \\
 & & XL & 2 & 1 & 8 & 4 & 38.059 & 13.606 & 16 &  6 & 39 & 10 & 765 & 832 & $<$0.001 & 0.48 (\cL) \\
\cmidrule{2-17}
 & \multirow{5}{*}{CI/CD jobs}
   & XS & 0 & 0 & 0 & 0 &  2.915 & 2.965 & 3 & 0 & 4 & 5 & 282 & 173 & $<$0.001 & 0.14 (\cN) \\
 & & S  & 0 & 0 & 0 & 0 &  4.623 & 4.200 & 3 & 0 & 6 & 6 & 186 & 211 & $<$0.001 & 0.13 (\cN) \\
 & & M  & 0 & 0 & 0 & 0 &  7.625 & 5.324 & 4 & 0 & 10 & 7 & 309 & 261 & $<$0.001 & 0.21 (\cS) \\
 & & L  & 0 & 0 & 0 & 0 & 14.234 & 6.884 & 7 & 0 & 17 & 6 & 530 & 372 & $<$0.001 & 0.30 (\cS) \\
 & & XL & 0 & 0 & 0 & 0 & 20.839 & 6.341 & 7 & 4 & 18 & 5 & 547 & 675 & $<$0.001 & 0.25 (\cS) \\

\midrule

\multirow{30}{*}[-60pt]{\rotatebox[origin=c]{90}{\textbf{ONAP}}}
 & \multirow{5}{*}{\#Revisions}
   & XS & 1 & 1 & 1 & 1 & 1.911 & 1.566 & 1 & 1 & 2 & 2 &  19 &  22 & $<$0.001 & 0.16 (\cS) \\
 & & S  & 1 & 1 & 1 & 1 & 2.432 & 2.025 & 2 & 1 & 3 & 2 &  27 &  47 & $<$0.001 & 0.19 (\cS) \\
 & & M  & 1 & 1 & 1 & 1 & 2.850 & 2.561 & 2 & 2 & 3 & 3 &  23 &  43 & $<$0.001 & 0.14 (\cN) \\
 & & L  & 1 & 1 & 1 & 1 & 4.011 & 4.077 & 2 & 2 & 4 & 5 &  77 & 103 &    0.018 & 0.04 (\cN) \\
 & & XL & 1 & 1 & 2 & 1 & 5.099 & 4.343 & 3 & 2 & 6 & 5 &  42 &  72 & $<$0.001 & 0.20 (\cS) \\
\cmidrule{2-17}
 & \multirow{5}{*}{\#Rebases}
   & XS & 0 & 0 & 0 & 0 & 0.567 & 0.181 & 0 & 0 & 1 & 0 &  10 &  17 & $<$0.001 & 0.25 (\cS) \\
 & & S  & 0 & 0 & 0 & 0 & 0.764 & 0.248 & 1 & 0 & 1 & 0 &  10 &  10 & $<$0.001 & 0.33 (\cM) \\
 & & M  & 0 & 0 & 0 & 0 & 0.802 & 0.327 & 1 & 0 & 1 & 0 &  12 &  12 & $<$0.001 & 0.34 (\cM) \\
 & & L  & 0 & 0 & 0 & 0 & 0.953 & 0.457 & 1 & 0 & 1 & 1 &  18 &  13 & $<$0.001 & 0.31 (\cS) \\
 & & XL & 0 & 0 & 0 & 0 & 1.081 & 0.503 & 1 & 0 & 1 & 0 &  12 &  23 & $<$0.001 & 0.39 (\cM) \\
\cmidrule{2-17}
 & \multirow{5}{*}{\makecell{Review dur.\\(hours)}}
   & XS & 0 & 0 & 1.9 &  2.1 & 107.354 &  91.794 & 15.056 & 18.035 &  71.062 &  71.606 & 4306.5 & 14296 &    0.331 & $-$0.02 (\cN) \\
 & & S  & 0 & 0 & 3 &  3.3 & 137.344 & 109.260 & 20.265 & 21.815 &  91.860 &  91.851 & 6031.2 & 14849 &    0.838 & $-$0.00 (\cN) \\
 & & M  & 0 & 0 & 3.3 &  4.1 & 125.871 & 136.358 & 25.111 & 24.938 & 118.099 & 117.314 & 4698.9 & 10559.6 &    0.603 & $-$0.01 (\cN) \\
 & & L  & 0 & 0 & 5.1 & 10.6 & 174.648 & 223.469 & 43.306 & 51.805 & 170.484 & 192.646 & 6353.3 & 10430.8 & $<$0.001 & $-$0.06 (\cN) \\
 & & XL & 0 & 0 & 7 &  4.7 & 267.763 & 191.963 & 70.475 & 38.814 & 261.642 & 167.277 & 5582.2 &  5708.4 & $<$0.001 & 0.12 (\cN) \\
\cmidrule{2-17}
 & \multirow{5}{*}{\makecell{Discussion\\messages}}
   & XS & 0 & 0 & 6 & 6 & 12.801 & 12.855 &  8 &  9 & 13 & 14 & 181 & 227 &    0.256 & $-$0.02 (\cN) \\
 & & S  & 0 & 0 & 7 & 7 & 15.255 & 16.322 &  9 &  9 & 15 & 17 & 235 & 405 &    0.388 & $-$0.02 (\cN) \\
 & & M  & 1 & 0 & 7 & 7 & 16.821 & 18.394 & 10 & 10 & 16 & 20 & 268 & 430 &    0.460 & $-$0.02 (\cN) \\
 & & L  & 0 & 0 & 7 & 7 & 24.363 & 27.364 & 11 & 13 & 21 & 31 & 748 & 530 & $<$0.001 & $-$0.07 (\cN) \\
 & & XL & 0 & 0 & 9 & 7 & 25.774 & 26.381 & 14 & 13 & 26 & 27 & 473 & 626 &    0.016 & 0.08 (\cN) \\
\cmidrule{2-17}
 & \multirow{5}{*}{\makecell{Inline\\comments}}
   & XS & 2 & 1 & 7 & 7 & 14.719 & 15.893 &  9 & 11 & 14 & 18 & 204 & 259 & $<$0.001 & $-$0.11 (\cN) \\
 & & S  & 2 & 2 & 8 & 8 & 17.544 & 20.147 & 10 & 12 & 17 & 22 & 250 & 454 & $<$0.001 & $-$0.10 (\cN) \\
 & & M  & 2 & 2 & 8 & 8 & 19.892 & 23.796 & 11 & 13 & 18 & 27 & 305 & 476 & $<$0.001 & $-$0.10 (\cN) \\
 & & L  & 2 & 2 & 8 & 9 & 29.261 & 35.821 & 13 & 17 & 26 & 41 & 820 & 968 & $<$0.001 & $-$0.12 (\cN) \\
 & & XL & 2 & 2 & 10 & 9 & 30.737 & 33.516 & 16 & 15 & 31 & 33 & 518 & 712 &    0.287 & 0.03 (\cN) \\
\cmidrule{2-17}
 & \multirow{5}{*}{CI/CD jobs}
   & XS & 0 & 0 & 0 & 0 & 1.918 & 3.038 & 2 & 2 & 3 & 3 &  39 &  64 & $<$0.001 & $-$0.22 (\cS) \\
 & & S  & 0 & 0 & 0 & 1 & 2.289 & 3.824 & 2 & 2 & 3 & 4 &  44 &  74 & $<$0.001 & $-$0.24 (\cS) \\
 & & M  & 0 & 0 & 0 & 2 & 3.071 & 5.402 & 2 & 3 & 4 & 6 & 107 & 108 & $<$0.001 & $-$0.26 (\cS) \\
 & & L  & 0 & 0 & 0 & 2 & 4.897 & 8.457 & 2 & 3 & 5 & 9 & 347 & 543 & $<$0.001 & $-$0.22 (\cS) \\
 & & XL & 0 & 0 & 0 & 2 & 4.964 & 7.135 & 1 & 2 & 6 & 7 & 186 & 164 & $<$0.001 & $-$0.14 (\cN) \\

\bottomrule
\end{tabular}
}
\end{table}

\textbf{Finding \#2.4: Middle-of-chain members function as synchronization bottlenecks within dependency-linked review workflows.}
Middle-of-chain members function as synchronization bottlenecks within dependency-linked review workflows. Across all three ecosystems, middle-of-chain changes show higher median \#Revisions, more \#Rebases, and longer review duration than the base or the top of the same chain (Table~\ref{tab:rq2-position-full}). The per-project view confirms the pattern: in nearly every project, \#Revisions medians at middle positions sit at or above those at the base and the top (Figure~\ref{fig:position-patchsets}). This matches bottom-up merge enforcement: a middle member is simultaneously blocked by its predecessor and blocking its successor, accumulating the iteration cost of waiting for changes below while continuing to receive feedback from above. As a result, middle members effectively function as synchronization bottlenecks that absorb the highest coordination burden within the chain. Reviewers internalise this dependency: on a 17-member ONAP \texttt{sdc} chain (change~88612\footnote{\url{https://gerrit.onap.org/r/c/sdc/+/88612}}), one held back approval with ``\emph{waiting for parent to be verified}.''

\begin{rqsummary}{2}
Relation chains introduce synchronization overhead through increased revisions, rebasing activity, review duration, and amplified CI workload compared to isolated changes (Finding~\#2.1). These coordination costs emerge primarily from dependency management rather than increased conversational review activity (Finding~\#2.2). Under strict CI-gating regimes, relation chains exhibit a \emph{CI amplification effect}, where revisions to one member repeatedly retrigger validation across dependent descendants (Finding~\#2.3). Within chains, middle members function as synchronization bottlenecks that absorb the highest coordination burden due to simultaneous upstream and downstream dependency constraints (Finding~\#2.4).
\end{rqsummary}

\subsection{RQ3: Structural Factors}
\label{sec:rq3}

\begin{figure}[bt!]
\centering
\includegraphics[width=\textwidth,height=5cm]{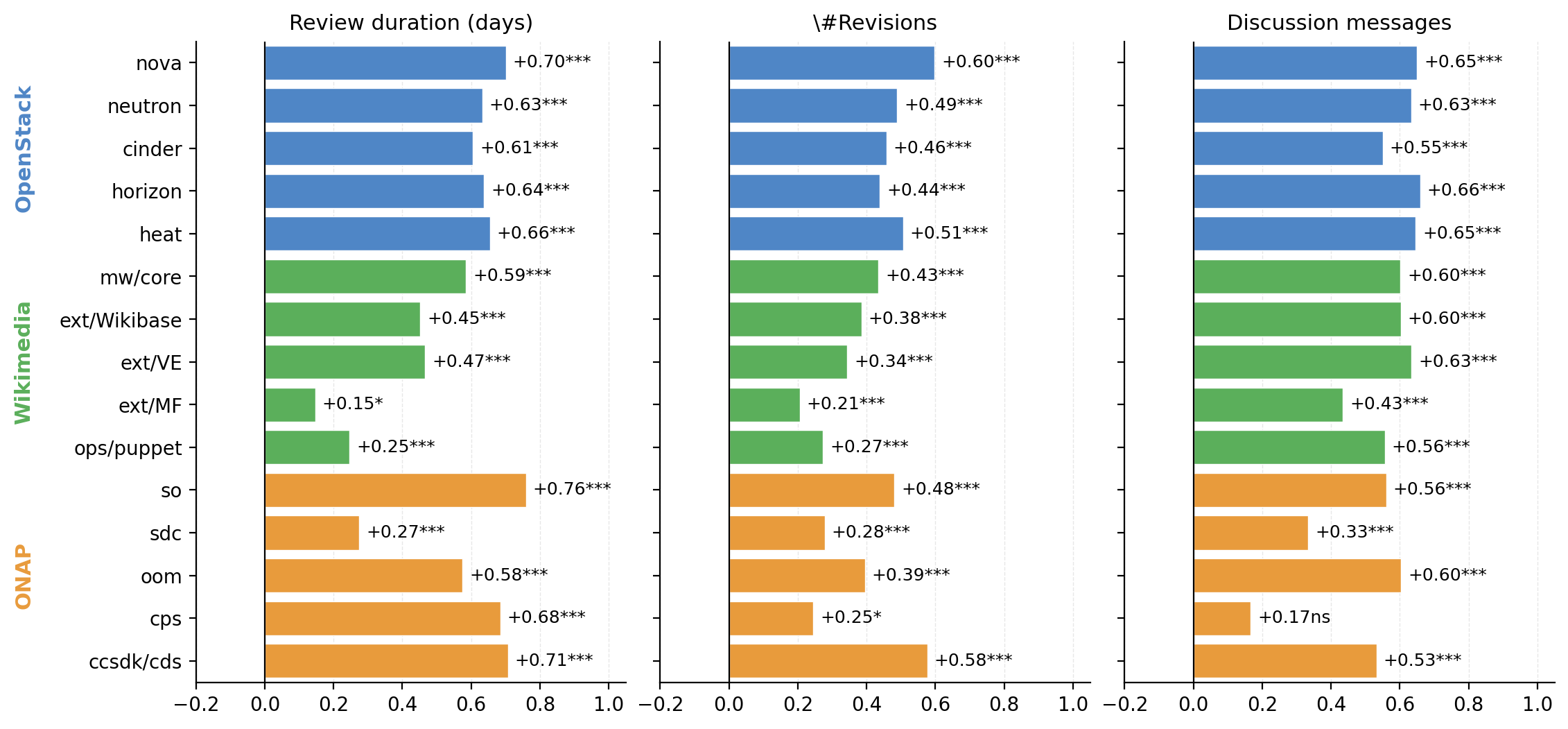}
\caption{Foundation effect: per-chain Spearman correlation between base-change and descendant-mean review metrics, computed within each project.}
\label{fig:foundation}
\end{figure}

Table~\ref{tab:chain-timing} reports submission and merge gaps between consecutive chain members. Figure~\ref{fig:foundation} shows Spearman correlations between base-change and descendant-mean review metrics.

\textbf{Finding \#3.1: Chain throughput is set by the project's CI gating regime rather than chain mechanics.}
Relation chains unfold in two timing regimes: a \emph{submission gap} (time between uploads of consecutive members) and a \emph{merge gap} (time between merges). Submission is fast on aggregate: pooled median 14 minutes, with project medians from near-zero in OpenStack \texttt{Horizon} and \texttt{Heat} to roughly six hours in ONAP \texttt{sdc} and \texttt{cps} (Table~\ref{tab:chain-timing}); these reflect developer work patterns, since a multi-step refactor prepared offline appears within minutes while incremental uploads widen the spread. Merge gaps separate ecosystems more sharply: medians sit below one hour for most Wikimedia and ONAP projects but between two and ten hours for every OpenStack project (\texttt{Horizon} 2.23~h to \texttt{Neutron} 9.82~h). The split tracks CI-gating regimes~\cite{maipradit2023repeated,cassee2020silent}: OpenStack runs a per-member test pipeline before each descendant's submission completes, while Wikimedia and ONAP permit descendants to follow within minutes once the base has merged. The Horizon \texttt{pylint} cleanup of December~2018\footnote{\url{https://review.opendev.org/q/topic:pylint+project:openstack/horizon}} shows this at chain level: chains assembled offline (small submission gaps) produced merge gaps in hours because each descendant waited for its predecessor's CI pipeline on the rebased revision. The same dependency structure therefore produces substantially different coordination costs under different CI-gating regimes.

\begin{table}[!htbp]
\centering\scriptsize
\caption{Per-project descriptive statistics of submission and merge gaps (in hours) between consecutive chain members. \texttt{ops/puppet} is excluded from aggregate statistics (see Table~\ref{tab:corpus}).}
\label{tab:chain-timing}
\renewcommand{\arraystretch}{1.05}
\resizebox{\textwidth}{!}{%
\begin{tabular}{l l r r r r r r r r r r r r}
\toprule
& & \multicolumn{6}{c}{\textbf{Submission gap (h)}} & \multicolumn{6}{c}{\textbf{Merge gap (h)}} \\
\cmidrule(lr){3-8}\cmidrule(lr){9-14}
\textbf{Ecosystem} & \textbf{Project}
 & \textbf{Min} & \textbf{Q1} & \textbf{Med} & \textbf{Mean} & \textbf{Q3} & \textbf{Max}
 & \textbf{Min} & \textbf{Q1} & \textbf{Med} & \textbf{Mean} & \textbf{Q3} & \textbf{Max} \\
\midrule
OpenStack  & nova         & 0.00 & 0.00  & 0.10 & 222.7 & 21.06 & 41{,}640
                          & 0.00 & 0.02  & 4.62  & 175.2 & 65.80 & 17{,}956 \\
           & neutron      & 0.00 & 0.00  & 0.56 & 267.6 & 74.75 & 22{,}172
                          & 0.00 & 0.06  & 9.82  & 124.1 & 72.69 & 24{,}983 \\
           & cinder       & 0.00 & 0.00  & 0.36 & 525.9 & 96.41 & 62{,}353
                          & 0.00 & 0.01  & 4.35  & 149.8 & 69.50 & 7{,}079 \\
           & horizon      & 0.00 & 0.00  & 0.00 & 455.6 & 35.78 & 14{,}003
                          & 0.00 & 0.04  & 2.23  & 129.5 & 70.95 & 2{,}520 \\
           & heat         & 0.00 & 0.00  & 0.00 & 268.6 & 18.12 & 9{,}691
                          & 0.00 & 0.06  & 4.81  & 263.2 & 72.62 & 19{,}390 \\
\midrule
Wikimedia  & mw/core      & 0.00 & 0.00  & 0.11 & 229.7 & 4.02  & 49{,}514
                          & 0.00 & 0.00  & 0.09  & 23.3  & 1.17  & 8{,}268 \\
           & ext/Wikibase & 0.00 & 0.00  & 0.80 & 104.5 & 21.32 & 38{,}206
                          & 0.00 & 0.01  & 0.27  & 22.5  & 2.91  & 7{,}255 \\
           & ext/VE       & 0.00 & 0.00  & 0.25 & 184.8 & 6.46  & 45{,}289
                          & 0.00 & 0.00  & 0.06  & 36.8  & 1.96  & 8{,}353 \\
           & ext/MF       & 0.00 & 0.00  & 0.31 & 72.3  & 21.07 & 4{,}635
                          & 0.00 & 0.03  & 0.53  & 18.9  & 8.11  & 601 \\
           & ops/puppet$^\dagger$ & 0.00 & 0.26 & 1.32 & 155.0 & 24.66 & 53{,}250
                          & 0.00 & 0.10  & 0.28  & 2.74  & 0.74  & 3{,}984 \\
\midrule
ONAP       & so           & 0.00 & 0.06  & 0.31 & 35.9  & 2.15  & 7{,}323
                          & 0.00 & 0.00  & 0.005 & 7.1   & 0.11  & 1{,}064 \\
           & sdc          & 0.00 & 0.45  & 5.83 & 84.9  & 71.23 & 3{,}474
                          & 0.00 & 0.04  & 0.63  & 6.7   & 3.31  & 555 \\
           & oom          & 0.00 & 0.00  & 0.21 & 61.8  & 18.26 & 2{,}809
                          & 0.00 & 0.00  & 0.007 & 31.8  & 1.05  & 1{,}372 \\
           & cps          & 0.00 & 0.29  & 5.83 & 90.9  & 100.79 & 959
                          & 0.00 & 0.19  & 0.65  & 14.3  & 2.40  & 311 \\
           & ccsdk/cds    & 0.00 & 0.02  & 0.98 & 37.6  & 17.48 & 1{,}468
                          & 0.00 & 0.00  & 0.006 & 5.4   & 0.56  & 242 \\
\bottomrule
\end{tabular}
}

\smallskip
{\scriptsize $^\dagger$Excluded from aggregate statistics (see Table~\ref{tab:corpus}).}
\end{table}

\textbf{Finding \#3.2: Higher review activity on base is associated with higher review activity on chain descendants.}
\label{sec:foundation}Spearman correlations between the base change's review metrics and the mean of its descendants' metrics are positive across all three metrics: review duration (median~$\rho = +0.61$, significant in 15/15~projects), discussion messages ($\rho = +0.60$, 14/15), and \#Revisions ($\rho = +0.43$, 15/15) (Figure~\ref{fig:foundation}). We refer to this regularity as the \emph{foundation effect}: chains with high-effort bases have high-effort descendants, and chains with low-effort bases have low-effort descendants. Because our metrics measure review activity rather than code quality, we describe the correlation as review-effort propagation, not evidence of base quality. The Nova SDK migration chain (change~659691)\footnote{\url{https://review.opendev.org/c/openstack/nova/+/659691}} illustrates the pattern: its base accumulated 38 \#Revisions and 17~discussion messages, and its 14 merged descendants each required additional review activity.

\vspace{-0.7cm}

\begin{rqsummary}{3}
The median merge gap between consecutive chain members varies by roughly three orders of magnitude across ecosystems and is explained by the project's CI-gating regime rather than chain mechanics (Finding~\#3.1). The \emph{foundation effect}: base-change review activity correlates positively with descendant review activity (Spearman~$\rho = 0.43$--$0.61$) in 14--15 of 15~projects (Finding~\#3.2).
\end{rqsummary}

\section{Discussions and Implications}
\label{sec:discussion}

Although many observed effect sizes are small according to Romano et al.'s Cliff's $\delta$ thresholds, the directional consistency of the results across projects and ecosystems suggests that relation-chain membership captures recurring coordination patterns rather than isolated project-specific anomalies.

\noindent \textbf{Implications for Researchers.} The change-level unit of analysis common in code-review research does not capture review dynamics in ecosystems where a substantial share of changes belong to chains; modern review workflows increasingly operate as dependency-aware coordination systems rather than collections of independent review units. When a chain is present, the review process involves decisions across dependent changes (which member to review first, whether to wait for a predecessor, how to handle merge ordering) invisible from any single change. Studies of reviewer load, reviewer assignment~\cite{thongtanunam2015whoreview,mirsaeedi2020mitigating}, and review latency~\cite{yu2015wait,zhang2022prdecision} that treat reviews independently may not represent the full process, and predictive models built on change-level features alone omit chain-level signals (chain position, depth, base-change activity, chain dynamism) that our data show correlate with the same outcomes they aim to predict.

Two further chain-level properties deserve attention. First, chains are not static dependency graphs but continuously evolve throughout review, a behaviour we refer to as \emph{dynamic dependency restructuring}: 33.5\% of chain members undergo a structural change in their parent SHA before merging, most often by detaching from an in-review parent (40\% of evolution events), with rates of 25.0\% (OpenStack), 39.7\% (Wikimedia), and 50.2\% (ONAP). Reviewers describe these dynamics in our metadata: on Wikimedia \texttt{mediawiki/core} change~530540\footnote{\url{https://gerrit.wikimedia.org/r/c/mediawiki/core/+/530540}}, one noted that ``\emph{the chain is now just blocked on the MediaWikiTestCaseBase thing},'' treating the chain as a unit gated by a single member. Second, chains persist across multi-year timespans: submission gaps reach $\sim$4.8~years in OpenStack \texttt{Nova} and merge gaps $\sim$2.85~years in \texttt{Neutron} (Table~\ref{tab:chain-timing}). Chain-aware tooling must continuously recompute dependency topology throughout the review lifecycle.

The foundation effect (Finding~\#3.2) shows base-change review activity correlates with descendant-change activity. Possible mechanisms include scrutiny applied to descendants of a heavily reviewed base, descendants inheriting architectural decisions, and reviewers accumulating context across the chain; our observational data does not distinguish between them, and because our metrics measure review activity rather than code correctness we do not interpret the correlation as evidence of base quality. LLM-based review agents~\cite{li2022automating,tufano2022using,guo2024exploring} process changes independently of chain dependency, yet chain position is associated with differences in review effort (Finding~\#2.4) and one third of chain members restructure during review; a natural extension is to provide such tools with chain-level inputs: position, base review state, and recent dependency-graph evolution. Chain prevalence, evolution rates, and median merge gaps also vary across ecosystems by amounts comparable to chain-vs-solo differences within a single ecosystem (Findings~\#1.2, \#3.1), correlating with submit-type and CI-gating configuration~\cite{maipradit2023repeated,cassee2020silent}; replications should evaluate chain practices per setting rather than pooling across ecosystems.

\noindent\textbf{Implications for Practitioners.} Our findings argue against the current default in review tooling, which processes each change as an independent unit~\cite{thongtanunam2015whoreview,mirsaeedi2020mitigating,li2022automating,tufano2022using}: when changes belong to a relation chain, their review outcomes are quantitatively linked. The foundation effect (Finding~\#3.2) is the central evidence: Spearman $\rho = 0.43$--$0.61$ in 14--15 of 15~projects show base review activity predicts descendant activity. The practical consequence is a single high-leverage intervention point per chain. Concentrating reviewer attention on the base (through earlier reviewer assignment, dedicated review slots, or AI-assisted summarisation of the base's open discussion) is associated with lower per-descendant review effort, multiplying the effect across every member. The Nova SDK migration chain (change~659691) illustrates the scale: 38 \#Revisions and 17 discussion messages on the base, with 14 merged descendants each absorbing additional review activity that would likely have been smaller had the base resolved with less friction.

Within chains, middle-of-chain changes show the highest median review effort (Finding~\#2.4), since a middle change is simultaneously blocked by its predecessor and blocking its successor. Review dashboards and reviewer-assignment systems~\cite{thongtanunam2015whoreview,mirsaeedi2020mitigating} could use chain position as an input, prioritising changes whose merge unblocks the largest number of downstream changes. Chain-aware CI could reorder test-resource scheduling. Consider the Horizon \texttt{pylint} cleanup of December~2018: its seventeen chains were assembled in one afternoon (Finding~\#3.1), but descendants queued behind each other for hours, each waiting for its predecessor's CI pipeline. A chain-aware scheduler could have given higher priority to the base (gating seventeen downstream pipelines) and lower priority to any single descendant (gating none until the base merges), turning hours of serialised wait into parallel queueing once the base passes. The principle generalises: OpenStack chain members run 10--23 CI/CD jobs per position against fewer than two for solo changes (Finding~\#2.3), so wasted resources on superseded patchsets are a recurring property of strict-gating regimes.

Finally, a chain's dependency graph is not fixed during review: 33.5\% of chain members restructure before merging, most often by detaching from an in-review parent. Tools that read the graph once and assume it stable (reviewer recommendation, CI scheduling heuristics, dashboards) produce results based on a graph no longer holding in about one-third of cases. A practical mitigation is to recompute chain-dependent state on every patchset upload, not only at chain creation; the long-tail timing in Table~\ref{tab:chain-timing} further suggests chain-aware tooling should preserve coordination state across dormant periods rather than discarding context after fixed timeouts.
\section{Threats to Validity}
\label{sec:threats}
\textbf{Construct validity.} We measure review effort (latency, patchset count, comment count) rather than review quality; the foundation effect is therefore framed as propagation of review activity, not of code quality. Chain detection relies on parent-SHA matching, validated against Gerrit's Relation Chain panel, with the operations/puppet auto-generated deep chain excluded from aggregates and chain evolution distinguishing true structural changes from trivial rebases (same Change-Id). Human-vs-bot distinction in discussion and CI/CD metrics relies on a per-project allowlist; while this convention is standard in Gerrit-based MCR research~\cite{mcintosh2016empirical,wessel2022quality}, residual misclassification at the long tail of commenters cannot be excluded. The Nova patchset-level data was extracted through a replication package rather than our own pipeline; we verified that the same Change-Id-based exclusion was applied, and removing Nova does not qualitatively change the results.

\noindent \textbf{Internal validity.} We compare chain members and solo changes within size strata but cannot exclude task complexity as an unmeasured confounder, so the latency overhead is treated as an association rather than a causal effect~\cite{arcuri2014hitchhiker}. We do not apply multiple-comparison corrections to per-project tests because projects are independent observational units, and the consistency of sign and magnitude across the corpus is the primary finding.

\noindent \textbf{External validity.} The corpus spans three open-source Gerrit ecosystems, dominated by infrastructure and platform software. Industrial settings, other platforms (GitHub~\cite{zhang2022prdecision}, Phabricator, Graphite), and different Gerrit submit types may exhibit different dynamics. Replication on those platforms is a natural next step.

\section{Conclusion}
\label{sec:conclusion}
We characterized 29{,}580 relation chains across 401{,}256 reviewed changes in three open-source Gerrit ecosystems. Chains are prevalent (5--49\% of changes, increasing in 14 of 15 projects) and impose measurable coordination costs: members take a median of 2.6$\times$ longer to merge than solo changes, exhibit a \emph{CI amplification effect} under strict gating, and place \emph{synchronization bottlenecks} on middle members. The \emph{foundation effect} (Spearman $\rho = 0.43$--$0.61$) shows base-change review activity propagates to descendants, and \emph{dynamic dependency restructuring} reshapes one third of chain members mid-review, sometimes across multi-year intermittent timespans.

These regularities suggest a lever for chain-aware tooling: invest review attention in base changes, and the benefit propagates through the chain. Future work should evaluate targeted interventions (base-review checkpoints, chain-aware reviewer assignment~\cite{thongtanunam2015whoreview,mirsaeedi2020mitigating}, dependency-graph-aware CI scheduling) and replicate the analyses on platforms beyond Gerrit (Phabricator, Graphite, GitHub stacked PRs).

\section{Data Availability}
\label{sec:data-availability}
To facilitate replication and extension, we share data and scripts in our replication package~\cite{replication_package}. The package contains raw Gerrit REST API outputs for the 15~studied projects, processed per-change and per-chain CSVs used as inputs to all analyses, and Jupyter notebooks that reproduce every figure and table reported in this paper, organised by research question. A README documents environment setup and provides a finding-to-cell mapping from each result in Sections~\ref{sec:rq1}--\ref{sec:rq3} to the specific notebook cell that produced it.

\bibliography{references}

\end{document}